\documentclass[conference,compsoc]{IEEEtran}
\usepackage{etoolbox}

\ifCLASSOPTIONcompsoc
  \usepackage[nocompress]{cite}
\else
  \usepackage{cite}
\fi
\ifCLASSINFOpdf
\else
\fi

\usepackage{makecell}

\usepackage{enumitem}
\usepackage{tikz}
\usepackage{mdframed}
\newmdtheoremenv{theo}{\textbf{Definition}}

\usepackage{amsmath}
\usepackage{etoolbox}

\usepackage{amssymb}
\usepackage{subfigure}
\usepackage[hidelinks]{hyperref}
\usepackage{cleveref}
\usepackage{xspace}
\usepackage{xcolor}

\usepackage{multirow}

\usepackage{algorithm}
\usepackage{algpseudocode}

\usepackage{tcolorbox}

\DeclareFontFamily{U}{mathx}{\hyphenchar\font45}
\DeclareFontShape{U}{mathx}{m}{n}{
      <5> <6> <7> <8> <9> <10>
      <10.95> <12> <14.4> <17.28> <20.74> <24.88>
      mathx10
      }{}
\DeclareSymbolFont{mathx}{U}{mathx}{m}{n}
\DeclareFontSubstitution{U}{mathx}{m}{n}
\DeclareMathAccent{\widecheck}{0}{mathx}{"71}
\def\etal{\emph{et al}.\xspace}

\def\DHdist{$\widecheck{H}$\xspace}

\newcommand\res[2]{{\small $#1$}{\tiny{\text{$\pm #2$}}}}

\newcommand{\ournameNoSpace}{\emph{RAGE}}

\newcommand{\ourname}{\ournameNoSpace\xspace}

\newcommand{\tracezero}{$\text{trace}_0$\xspace}

\algrenewcommand\algorithmicrequire{\textbf{Input:}}
\algrenewcommand\algorithmicensure{\textbf{Output:}}

\def\cross{$\times$\xspace}
\def\NA{N.A.\xspace}

\usepackage{hhline}
\usepackage{booktabs}

\newcommand{\specialcell}[2][c]{%
  \begin{tabular}[#1]{@{}c@{}}#2\end{tabular}}

\newtcolorbox{insight}[1]{colback=gray!5!white, colframe=gray!75!black, fonttitle=\bfseries, title={#1}, arc=0mm, boxrule=0.1mm}

\usepackage{titlesec}
\titlespacing*{\section}{0pt}{3.5pt}{3.5pt}
\titlespacing*{\subsection}{0pt}{3pt}{3pt}
\titlespacing*{\subsubsection}{0pt}{3pt}{3pt}
\usepackage[para]{footmisc}
\usepackage{comment}

\begin{document}
\title{\Large \bf One for All and All for One:\\GNN-based Control-Flow Attestation for Embedded Devices}

\author{
\IEEEauthorblockN{Marco Chilese\IEEEauthorrefmark{1},  Richard Mitev\IEEEauthorrefmark{1}, Meni Orenbach\IEEEauthorrefmark{2},\\ Robert Thorburn\IEEEauthorrefmark{3}, Ahmad Atamli\IEEEauthorrefmark{2}\IEEEauthorrefmark{3}, Ahmad-Reza Sadeghi\IEEEauthorrefmark{1}} \vspace{5px} 
\IEEEauthorblockA{\IEEEauthorrefmark{1}Technical University of Darmstadt, \IEEEauthorrefmark{2}NVIDIA, \IEEEauthorrefmark{3} University of Southampton}
}%\\

\maketitle

\begin{abstract}
Control-Flow Attestation (CFA) is a security service that allows an entity (verifier) to verify the integrity of code execution on a remote computer system (prover). Existing CFA schemes suffer from impractical assumptions, such as requiring access to the prover's internal state (e.g., memory or code), the complete Control-Flow Graph (CFG) of the prover's software, large sets of measurements, or tailor-made hardware. Moreover, current CFA schemes are inadequate for attesting embedded systems due to their high computational overhead and resource usage.

In this paper, we overcome the limitations of existing CFA schemes for embedded devices by introducing \ourname, a novel, lightweight CFA approach with minimal requirements. \ourname can detect Code Reuse Attacks (CRA), including control- and non-control-data attacks. It efficiently extracts features from \emph{one} execution trace and leverages Unsupervised Graph Neural Networks (GNNs) to identify deviations from benign executions. The core intuition behind \ourname is to exploit the correspondence between execution trace, execution graph, and execution embeddings to eliminate the unrealistic requirement of having access to a complete CFG.

We evaluate \ourname on embedded benchmarks and demonstrate that (i) it detects 40 real-world attacks on embedded software; (ii) Further, we stress our scheme with synthetic return-oriented programming (ROP) and data-oriented programming (DOP) attacks on the real-world embedded software benchmark Embench, achieving 98.03\% (ROP) and 91.01\% (DOP) F1-Score while maintaining a low False Positive Rate of 3.19\%; (iii) Additionally, we evaluate \ourname on OpenSSL, used by millions of devices and achieve 97.49\% and 84.42\% F1-Score for ROP and DOP attack detection, with an FPR of 5.47\%. 

\end{abstract}

\IEEEpeerreviewmaketitle

%-------------------------------------------------------------------------------
% SECTIONS
%-------------------------------------------------------------------------------
\section{Introduction}\label{sec:intro}

Remote attestation is a security service allowing an entity (verifier) to verify the integrity (and authenticity) of the software's state on a remote computing system (prover). Conventional static remote attestation schemes can only detect attacks that manipulate the binaries~\cite{seshadri2004swatt, seshadri2006scuba, chen2017secure, seshadri2005pioneer, li2011viper, seshadri2008sake, krauss2007detecting, agrawal2015program}, are unable to detect code-reuse attacks (CRAs), such as control-data attacks and non-control-data attacks (e.g., return-oriented programming attacks and data-oriented programming attacks, respectively). Yet, these attacks have increased significantly recently, especially on embedded devices; examples are the remote execution vulnerability on

\begin{figure}[ht]
	\centering
	\includegraphics[scale=0.6,trim={0 .62em 0 0},clip]{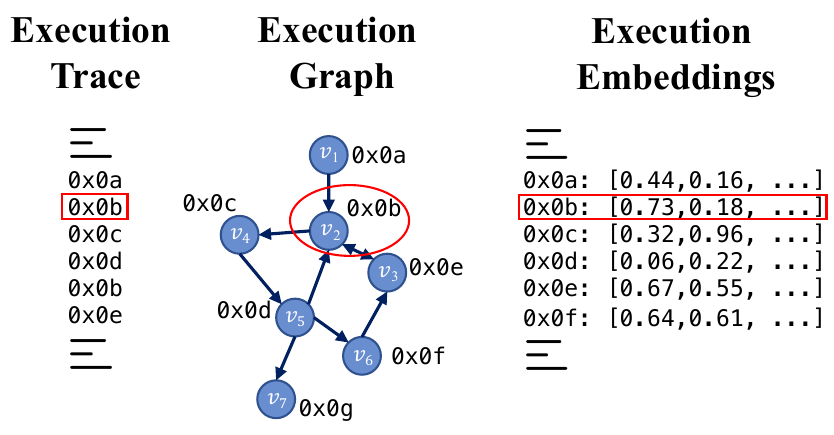}
        \caption{Correspondence between the recorded execution trace, the processed execution graph, and the execution embeddings produced by \ourname's model. The preserved mapping between basic blocks, graph nodes and embedding is highlighted in red.}
	\label{fig:exe_graph_emb}
\end{figure}
 Belkin's IoT smart plugs\footnote{CVE-2018-6692.}~\cite{belkin}, a remote access vulnerability to 200 million cable modems\footnote{CVE-2019-19494, CVE-2019-19495}~\cite{modems}, a privilege escalation vulnerability in Linux firewall\footnote{CVE-2022-25636}~\cite{firewall}, and allowing malware to bypass Samsung's phones ASLR (Address Space Layout Randomization)\footnote{CVE-2023-21492, CVE-2023-2149}~\cite{samsung}.

\textbf{Related work limitations.}
Control-Flow-Attestation (CFA)~\cite{abera2016c} schemes were proposed to overcome the limitations of static attestation. 
However, current CFA schemes suffer from several shortcomings: They typically assume the availability of a large set of execution measurements~\cite{abera2016c, zeitouni2017atrium, dessouky2017fat, dessouky2018litehax, sun2020oat, zhang2021recfa, yadav2023whole}, a complete Control-Flow Graph (CFG)~\cite{abera2016c, zeitouni2017atrium, dessouky2017fat, dessouky2018litehax, sun2020oat, zhang2021recfa, nunes2021tiny, yadav2023whole}, or specialized hardware~\cite{zeitouni2017atrium, dessouky2017fat, dessouky2018litehax, nunes2021tiny}.
The efficacy of approaches that require a full CFG directly relates to the completeness of the used CFG, which can only be approximated in practice~\cite{conrado2023bounded, frassetto2022cfinsight, theiling2002control, van2007relating, xu2009constructing, zhu2021constructing}. Although novel CFG extraction approaches improved on the approximation~\cite{zhu2021constructing}, extracting a complete CFG remains an open problem.
Further, existing attestation approaches based on machine learning have limitations, such as reliance on a-priori measurements and the need for prior knowledge of the system's internal state (e.g., memory, code)~\cite{hu2019probability, ma2019combination, aman2022machine}. Moreover, specialized hardware is not feasible for off-the-shelf devices. 

\textbf{Research Questions and Our Proposal.} 
In order to overcome the shortcomings of previous works, we aim to answer the following research questions:
\begin{description}[leftmargin=6pt]
    \item[(RQ1)] Is it possible to relax the assumption made by previous works about the availability of the complete CFG?
    \item[(RQ2)] Is it possible to preserve the inner characteristics of an execution trace while transposing it to a lower dimensional space (i.e., embeddings), leveraging a machine learning (ML) algorithm?
    \item[(RQ3)] Are the information extracted through a ML algorithm (i.e., execution embeddings) fine grained enough to separate benign executions from malicious ones?
    \item[(RQ4)] Are the execution embedding changes (i.e., structure-wise or behaviour-wise) related to code reuse attacks?
\end{description}

For thoroughly answering the research questions above, we designed \ourname, a novel CFA scheme that addresses the shortcomings of the existing CFA approaches; it does not require access to the complete CFG, a wide set of measurements, or specialized hardware. Our approach only needs a minimal set, basically only \emph{one}, of basic blocks' execution traces, not imposing any assumptions on the trace's completeness (coverage), or disclosing the memory content. Moreover, due to its low overhead, \ourname is adequate for deployment on embedded systems.

We overcome the limitations of previous works with an efficient approach that traces the program execution at the basic block level and utilizes an Unsupervised Graph Neural Network (GNN) to discern between partial Control-Flow Graphs of benign and malicious program executions. The primary rationale behind \ourname is to exploit the subtle relation between an execution trace, execution graph, and execution embedding, as shown in \Cref{fig:exe_graph_emb}. This approach results in a compact model which allows swift attestation. 
Similarly to works which assume an (approximated) complete CFG, \ourname works on incomplete execution graphs; more concretely, it utilizes only \emph{one} execution trace. However, by leveraging a deep unsupervised geometric model --- which learns the benign behavior of a program --- we can generalize over benign execution paths and identify control-flow deviations (i.e., control-data and non-control data attacks). We discuss this further in \Cref{sec:sec}.

Our evaluation focuses on various embedded software, which we evaluated with both real-world runtime attacks and synthetic attacks. 
We evaluated a set of 40 real-world attacks on embedded software, including return-to-libc and ROP attacks from a version of RIPE\footnote{\url{github.com/hrosier/ripe64}} framework~\cite{wilander.ripe} extended by our industry partner. The results demonstrate precise detection in a real-world scenario.

Since critical cryptographic operations typically rely on standard libraries, we also evaluated OpenSSL\footnote{\url{wiki.openssl.org/index.php/Libcrypto\_API}}, which was recently subject to remote-code-execution (RCE) attacks\footnote{CVE-2022-2274, CVE-2022-2068} and non-control-data attacks\footnote{CVE-2014-0160}. To test the capabilities of \ourname, we synthetically generate generic and stealthy ROP and DOP attacks.

Finally, aligned with the state-of-the-art~\cite{yadav2023whole}, we evaluated a popular and representative embedded benchmark suite Embench~\cite{bennett2022embench} that comprises 18 software packages, through the use of synthetic ROP and DOP attacks.

\textbf{Our Contributions.}
In summary our contributions include the following:
\begin{itemize}[leftmargin=*]
\setlength\parskip{-0.20em}
\item We introduce \ourname, a novel, lightweight CFA run-time attestation mechanism based on unsupervised Graph Neural Networks (GNNs). It does not require the program's Complete Control-Flow Graph, memory content, code, or a-priori measurements, preventing information leakage to the verifier. Our approach has low computational complexity in extracting features from an execution trace, making it particularly suitable for resource-constrained embedded devices. 

\item We show that Variational Graph Autoencoders (VGAEs) are suitable for extracting rich graph embeddings that preserve the control-flow behavior of the attested program. Our scheme's model can be trained effectively using just \emph{one} execution trace. To achieve this, \ourname finds the correspondence between execution trace's basic blocks, execution graph's nodes, and execution embeddings. This allows \ourname to eliminate the need for a complete control-flow graph compared to prior work.

\item We show that \ourname recognizes real-world attacks, such as ret2libc and return-oriented programming (ROP) attacks. Further, \ourname demonstrates a high level of accuracy in detecting synthetic code-reuse attacks, including ROP and data-oriented programming (DOP) attacks, on a popular representative embedded benchmark suite, with a F1-Score of 8.03\% (ROP) and 91.01\% (DOP) while maintaining a low 3.19\% FPR. Additionally, we evaluate such synthetic attack on cryptographic libraries, achieving 97.49\% (ROP) and 84.42\% (DOP) with a FPR 5.47\%.

\end{itemize}

\section{Background}\label{sec:background}
This section provides the background information necessary for understanding our work.

\subsection{Control-Flow Graph}\label{background:cfg}
A generic Control-Flow Graph (CFG) is the graph representation of the steps traversed during a program execution. More formally:
\begin{theo}
Given a program $P$, a $CFG_P$ is a directed graph $G=(V,E)$, where $V$ is the set of statements (i.e., nodes) and $E \subseteq V \times V$ is the set of edges. A control flow edge from the statement $s_i$ to $s_j$ is $e=(s_i, s_j) \in E$.
\end{theo}
Further, it is necessary to distinguish between a complete CFG (CCFG) and a partial CFG (PCFG):
\begin{theo}
A partial control-flow graph (PCFG) is the control flow graph of a program $P$, executed with input $i \in I$, where $I$ is the set of all the possible inputs. It is defined as: 
$$PCFG_{P(i)} = CFG_{P(i)}(V', E') \text{ s.t. } V' \subseteq V\text{, } E' \subseteq E\text{,}$$
where $V$ and $E$ are the nodes and edges set of $CCFG_P$.
\end{theo}

\begin{theo}\label{eq:ccfg}
A complete control-flow graph (CCFG) is the control flow graph of a program $P$, which includes all the possible paths in $P$ (i.e., for all the possible inputs). It is defined as:
$$CCFG_P = \bigcup_{i \in I} PCFG_{P(i)}\text{.}$$
\end{theo}

Intuitively, given the non-finite nature of the input set $I$, and considering the halting-problem~\cite{turing1936computable}, practically, it is not always possible to reconstruct a CCFG, as it falls under the family of NP-hard problems~\cite{granata2013maximum}. 
In real-world, CCFGs can only be approximated~\cite{conrado2023bounded, frassetto2022cfinsight, theiling2002control, van2007relating, xu2009constructing, zhu2021constructing}. More precisely, it is possible to build a CCFG \emph{only if a program $P$ does not contain} any indirect jump, any indirect branch, or any indirect call. Otherwise, it is impossible to ensure that all execution paths have been discovered~\cite{rimsa2021practical}. 

Recent works have shown that approximation approaches are still far from generating near-to-complete CFGs. Specifically, Rimsa~\textit{et al.}~\cite{rimsa2021practical} showed that static and dynamic analyses can be combined to generate more comprehensive CFGs. However, they can only approximate up to 46\% of SPEC CPU2017\footnote{\url{www.spec.org/cpu2017/}}. Zhu~\textit{et al.}~\cite{zhu2021constructing} reported that previous approximation approaches missed, on average 34.9\% of basic blocks. Moreover, approximation techniques are computationally intensive, requiring multicore CPUs, large amounts of memory, and several hours of computation, which are impractical, especially for resource-constrained devices.

\subsection{Control-Flow Attestation}
Control-Flow Attestation (CFA) is a critical security service to remotely verify the execution path integrity of a program~\cite{abera2016c}. 
CFA involves creating a record of the instructions sequence executed during the program's run-time and then using this record to verify that the program was executed correctly. This method can detect run-time software attacks or data input manipulation. CFA schemes have been proposed as solutions in software, hardware, or a combination of both (cf., \Cref{sec:related}).

\subsection{Code Reuse Attacks}\label{subsec:background_memory}
Code-Reuse Attacks (CRAs) are a type of run-time attack that hijack the control flow of a program by using the code segments already present in the program's memory. These attacks construct malicious code sequences, called gadgets, by chaining them in a specific order, enabling them to perform unauthorized actions such as program interruption, data tampering, exfiltrate sensitive information, or code execution. Remote-Code-Execution (RCE) attacks are a subset of CRAs that have been growing in number and complexity, accounting for 25\% of all reported vulnerabilities\footnote{\url{cvedetails.com/vulnerabilities-by-types.php}}.
The most common type of CRA is return-oriented programming (ROP)~\cite{shacham2007geometry}, where the attacker uses gadgets chained together through return instructions. In the past, many variations of ROP attacks were published, including Jump-oriented programming (JOP)~\cite{bletsch2011jump} or Load-oriented programming (LOP)~\cite{checkoway2010return}, which all work similarly, only distinguished by the type of instructions to link the gadgets.

Another type of CRA is non-control-data attacks~\cite{ispoglou2018block}, generalized as Data-oriented programming (DOP)~\cite{hu2016data} where the adversary manipulates the data stored in memory to change the flow of execution.

ROP, JOP, LOP, and DOP attacks change the benign control-flow of a program. However, DOP does not introduce new edges in the program's Control-Flow Graph (i.e., new jumps), and are generally harder to detect. 

Since the effect of ROP, JOP, and LOP attacks on the CFG are very similar, we continue referencing this family of control-data attacks as ROP attacks throughout this paper.

\subsection{Graph Neural Networks}\label{subsec:background_vgaes}
 Graph Neural Networks (GNNs) are types of Neural Networks (NN) that work with graph-structured data~\cite{sperduti1997supervised, welling2016semi}. 

Specifically, GNNs are used as a component of Graph Autoencoders (GAEs). GAEs are an unsupervised application of GNNs that transform graphs into a low-dimensional vector representation and then reconstruct the original graph data from that latent representation (i.e., a set of features that captures the underlying structure of data). Their usage includes compression and encoding graph data and feature extraction, a crucial task for graph generation in training.

\subsubsection{Variational Graph Autoencoders (VGAEs)}\label{subsec:vgae}

VGAEs are a type of GAEs introduced by Kipf \etal~\cite{kipf2016variational}, which are inspired by traditional Variational Autoencoders (VAEs)~\cite{kingma2013auto}. VGAEs combine the concepts of GAEs and VAEs. Like GAEs, they encode nodes and graphs into a low-dimensional vector representation (i.e., embedding or encoding) and then reconstruct the original graph data from that representation.

These types of models are trained by optimizing the variational lower bound $\mathcal{L}$ to approximate the true distribution of the training data:
\begin{equation}\label{eq:variational_low_bound}
    \mathcal{L} = \mathbb{E}_{q(Z|X,A)} \left[ \log{p(A|Z)} \right] - \text{KL} \left[ q(Z|X,A) \parallel p(Z) \right]
\end{equation}
  In the above equation, $X$ represents the features matrix of the nodes in the input graph, and $A$ is the graph's structure, such as the adjacency matrix. On the other hand, $q(Z|X,A)$ represents the inference model, which is defined as:
\begin{equation}\label{eq:vgae_inference}
\begin{split}
q(Z|X,A) = \prod_{i=1}^{N} q(z_i|X, A) \text{, with }\\
q(z_i|X,A)=\mathcal{N}(z_i|\mu_i, \text{diag}(\sigma^2_i)) 
\end{split}
\end{equation}
where $\mu$ and $\sigma$ are the output of graph convolutional layers (ConvGNN).
$p(Z)$ depicts the Gaussian prior \mbox{$p(Z) = \prod_{i=1}^N p(z_i)$}, where $z_i$ is the i-th element of the latent space array captured in $Z$. KL, instead, is the Kullback-Leibler divergence. Considering two probability distributions $Q$ and $P$ sampled from $\mathcal{X}$, KL is defined as the relative entropy, or the difference, from $Q$ to $P$, that is:
\begin{equation}\label{vgae:sampling}
    \text{KL}(P\parallel Q)=\sum_{x \in \mathcal{X}}P(x)\log \left({\frac{P(x)}{Q(x)}}\right) .
\end{equation}
After the training process, the VGAE's encoder is used to obtain graph embeddings.

\subsubsection{Graph Embeddings}
To extract crucial, yet not accessible, information from a graph $G$, we leverage the capability of VGAEs to extract latent features from the data. The process of graph embedding can be defined as follows.

\begin{theo}\label{def:embs}
    Considering a graph embedding model $f$ and a graph $G = (V, E)$, where V is the set of nodes and E is the set of edges, we define $Z_G$ as the embeddings of $G$. 
    $Z$ is the mapping of the original high-dimensional irregular domain (i.e., the graph) to a latent low-dimensional space, such that it is dense and continuous:
    $$Z_G \in \mathbb{R}^{|V|\times L} \text{ s.t. } L \ll |V|.$$

    The embedding model $f$ can be defined as: 
    $$f: v_i \rightarrow z_i \in \mathbb{R}^L$$
    where $v_i$ represents a node in the original space and $z_i$ represents its corresponding embedding in the low-dimensional space, establishing so a direct link between the two spaces.
\end{theo}

It should be noted that the node's properties, such as their structure, are preserved in the latent space, meaning that nodes close in the original space will also be close to each other in $Z$~\cite{xu2021understanding}. 
In our case, we utilize the VGAE model to perform the embedding process, resulting in an array $Emb_G \in \mathbb{R}^{n\times L}$ for a given graph $G$, where $n$ is the number of nodes in $G$.

\section{System and Threat Model}\label{sec:threat}

Our system model comprises two primary entities, an untrusted \emph{prover} device and a trusted \emph{verifier} device. The \emph{prover} device is equipped with a root of trust, such as a Trusted Execution Environment (TEE)\footnote{In practice, TEEs exist on of-the-shelf devices such as TrustZone or TrustZone-M, AMD SEV, Intel's SGX/TDX.}, which runs the tracing program and ensures its integrity (cf., \Cref{sec:impl:data_collection}) that cannot be compromised. 
Ultimately, an adversary can tamper with the software executed in the prover, e.g., by performing CRAs, but cannot manipulate the tracer, protected by the TEE, which ensures the integrity of the collected execution traces.

As we describe later our design in \Cref{sec:design}, \ourname requires a minimal number of traces for its operation. More concretely, it utilizes \emph{one} known benign trace of the program's execution for training the model and, at attestation time, as a reference, which we call \tracezero. The rest of the trace set is called the validation set, used for defining the attestation threshold (cf., \Cref{subsec:implementation_threshold}).
Subsequently, during each attestation request, the \emph{verifier} obtains a single execution trace to verify against \tracezero. On every update of the software to be attested, the \emph{verifier} will receive \emph{one} execution trace (i.e., new \tracezero) to retrain its model. Note that \ourname does not impose any assumptions about the path completeness of the generated (partial) CFG from \tracezero (i.e., assuming that the execution graph is falling under the definition of a PCFG).

We base the general design of the remote attestation scheme on already established mechanisms~\cite{abera2016c, yadav2023whole}. 
Analogously to related works~\cite{abera2016c, yadav2023whole}, we deem the performance of the tracer and its state-of-the-art implementation beyond the scope of this paper. With this, our approach is independent of the specific implementation of the tracer (either in software or in hardware) and existing tools can be used, as described in more detail in \Cref{sec:impl:data_collection}. Therefore, we aim to reduce the computational and storage overhead on the \emph{prover's} and \emph{verifier's} device.

\section{\ourname Design} \label{sec:design}
\begin{figure*}[!ht]
	\centering
	\includegraphics[scale=0.46,trim={0 0 0 0},clip]{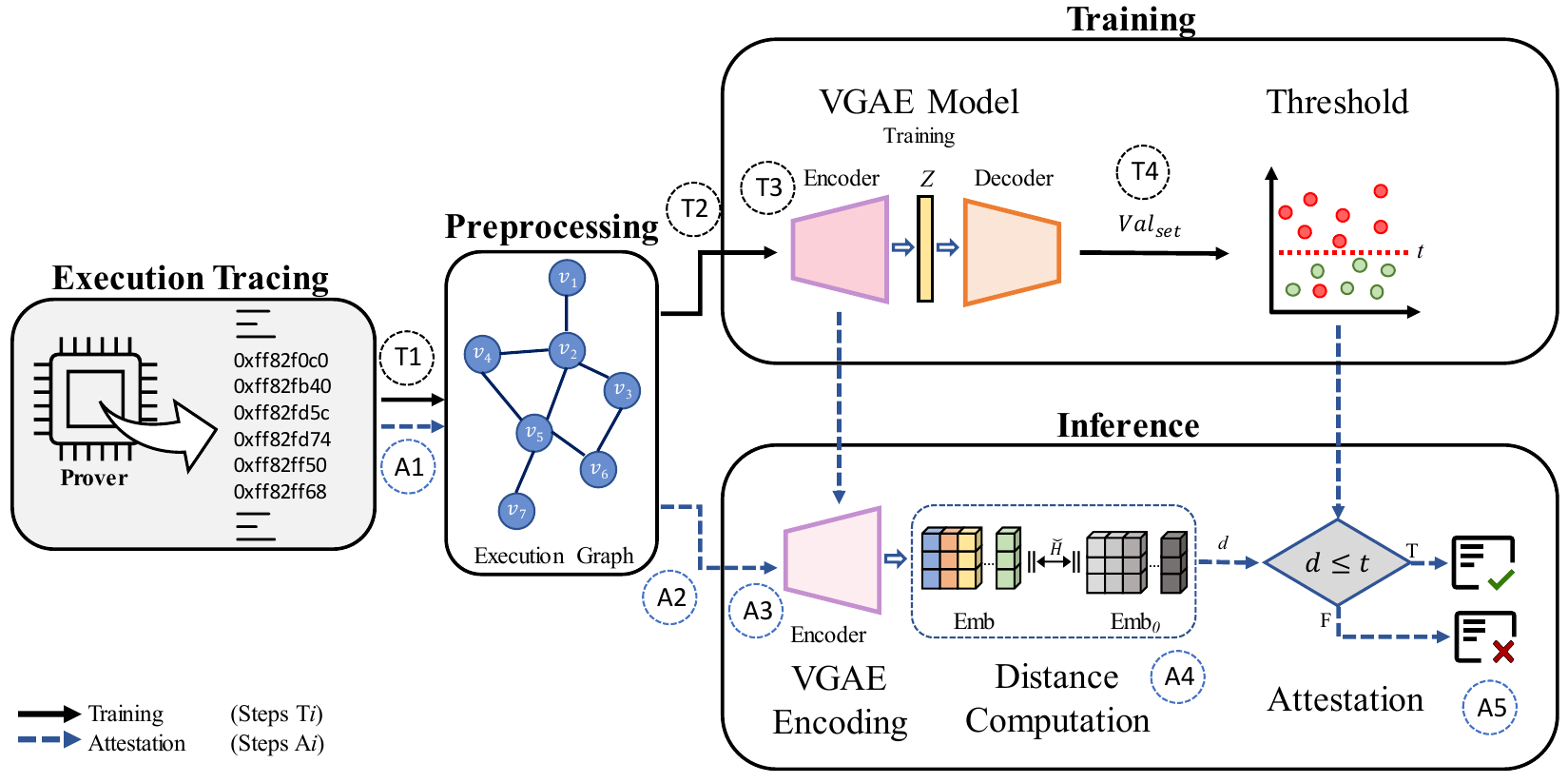}
        \caption{Overview of \ourname, including both the \emph{Training} pipeline (carried out by the verifier) and the \emph{Inference} stage (when attestation occurs).}
	\label{fig:sys_overview}
\end{figure*}

\ourname is a novel Control-Flow Attestation (CFA) scheme, the first approach leveraging Graph Neural Networks to accomplish run-time attestation. The VGAE model is a powerful technique for learning data that can be represented as a graph (i.e., graph-structured data). It can learn a compact representation of the execution graph for identifying deviations from benign patterns. This feature makes VGAE an ideal tool for anomaly detection in graph-structured data. 

We observed that unsupervised Graph Neural Networks and CFGs have a natural link. More precisely, the model learns the low-dimensional embeddings of the execution graphs (cf., \Cref{fig:exe_graph_emb}), which capture the underlying structure and patterns. Therefore, by combining these concepts, we can discern between benign and malicious execution graphs without the need for a CCFG.
As shown in \Cref{fig:sys_overview}, our scheme consists of a (i) training phase (solid black arrow) and (ii) attestation phase (dashed blue arrow).

\paragraph{\textbf{Training Phase}}
The prover starts collecting a set of known benign program execution traces (\textbf{T1}). The tracer only collects a list of traversed Program Counter (PC) basic block, which provides the information needed for reconstructing the execution flow. At the same time this limited data collection prevents information leakage to the verifier~\cite{molnar2006program} (cf.,~\Cref{sec:sec_completeness}). The collected addresses are used in two steps: \emph{one} trace is used for training the VGAE model and the remaining for calibrating the attestation threshold (cf., \Cref{subsec:implementation_threshold}).
Initially, the prover transmits the traces securely to the verifier. The verifier preprocesses the traces into an execution graph representation (\textbf{T2}), i.e., a PCFG, while extracting descriptive features of the control flow, maintaining a natural correspondence between basic block addresses and graph nodes, as depicted in \Cref{fig:exe_graph_emb}.

Note that the verifier does not need any knowledge about the code. Further, the preprocessing can also occur on the prover side, both in software or, if available, on specialized hardware to minimize data transmission (i.e., as shown in \Cref{{subsec:eval_runtime}}, network overhead can be minimized when representing traces directly as graphs).

Then, the verifier trains the VGAE model (\textbf{T3}) by optimizing the variational lower bound (cf., \Cref{subsec:vgae}). With the trained VGAE model's encoder, the verifier extracts embeddings (cf., Definition \ref{def:embs}) from the remaining trace set and calibrates the attestation threshold (\textbf{T4}).
The training phase can also occur offline. The embedded software provider may be a third party, separate from the vendor. It can also supply the attestation model and the threshold to the vendor who carries out the attestation. This removes the need to reveal the initial traces to the vendor.

\paragraph{\textbf{Attestation Phase}}
Now, once the training phase is completed, the attestation protocol will run. Upon receiving the attestation challenge, the prover traces one execution (\textbf{A1}), and preprocessing occurs (\textbf{A2}). Afterward, the verifier extracts embeddings  of the execution graph through the previously trained VGAE model's encoder (\textbf{A3}) and measures the embedding distance between the execution to attest and the training reference execution (\textbf{A4}) through the Directed Hausdorff distance (cf., \Cref{graph_dist}), which measures the (dis)similarity among the executions. Finally, the output of the threshold test is used to determine the attestation outcome (\textbf{A5}).

\ourname can detect unknown execution flow anomalies without needing a reference CCFG or its execution measurements (e.g., memory states), as it determines whether the execution is benign through the analysis of the embeddings' (dis)similarities.

\subsection{Geometric Deep Learning Model}\label{subsec:sys_overview_model}

\begin{figure*}[ht!]
    \centering
    \subfigure[Variational Graph Autoencoder (VGAE).]{\includegraphics[width=0.6\linewidth,clip]{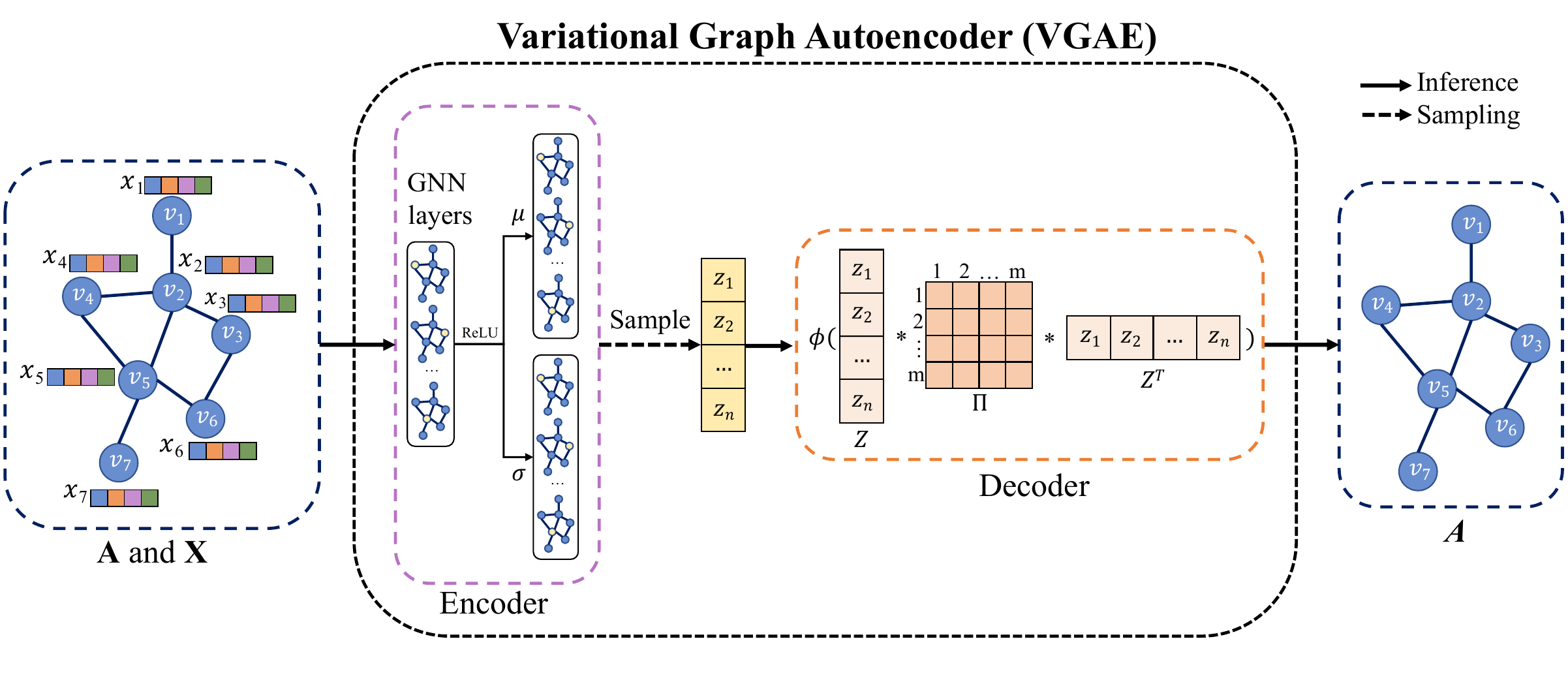}}
    %trim={0.2em 1em 0.2em 0}
    \hspace{1em}
    \subfigure[Our designed VGAE encoder.]{\includegraphics[width=0.31\linewidth,clip]{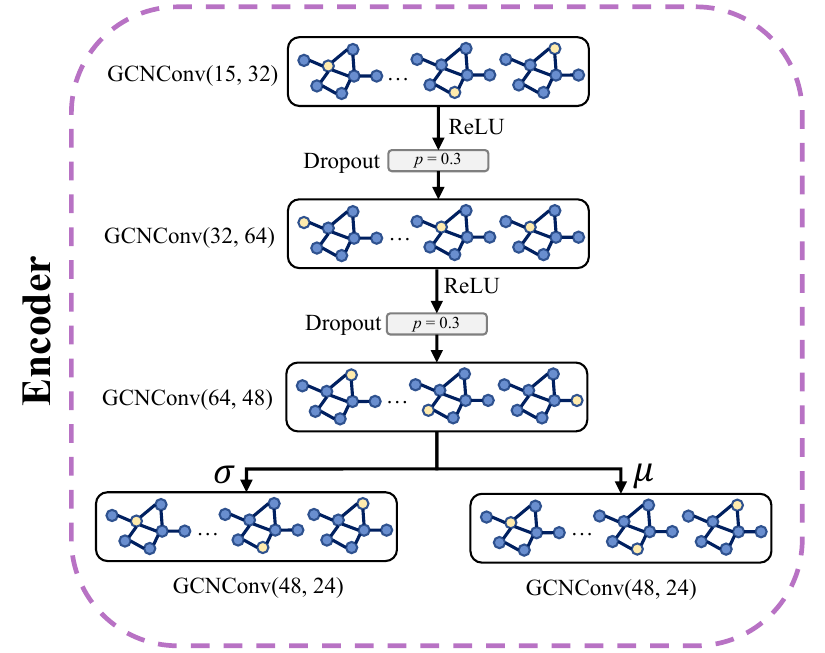}} 
    %trim={0 0 0 0}

    \caption{(a) Graphical representation of the Variational Graph AutoEncoders (VGAEs) structure. In the figure, the graph structure (i.e., graph connectivity in coordinate format - COO) is denoted as $A$, the features of each node are $X$, and $Z$ represents the latent variables. The operator $*$ represents the matrix product and $\phi(\cdot, \cdot)$ is the activation function. (b) Representation of the encoder designed for the Variational Graph Autoencoder (VGAE) utilized in \ourname. The decoder is not represented (i.e., the standard inner product decoder~\cite{kipf2016variational}). The overall number of parameters for the entire model is $8,128$.}
    \label{fig:VGAE} 
    \label{fig:VGAE_encoder}
\end{figure*}

After considering numerous deep learning approaches and models, we chose to adopt Graph Neural Networks, specifically a Variational Graph Autoencoder (VGAE) as the model of choice in the design of \ourname, as depicted in \Cref{fig:VGAE}a. The use of Graph Neural Networks is a novel and natural association with programs' execution traces, as these kinds of data can naturally be represented as a graph. The VGAE model is a powerful tool for graph representation learning that combines the strengths of both Graph Autoencoders (GAE) and Variational Autoencoders (VAE). GAEs learn to encode graph structures into a lower-dimensional latent space (i.e., embedding), while VAEs introduce a probabilistic approach to generate new graphs by sampling from a prior distribution (cf., \Cref{vgae:sampling}). By integrating these two frameworks, VGAEs generate new graphs and provide a probabilistic encoding of the input graph.

The encoder component of the VGAE is essential for this purpose, as it is responsible for mapping the input graph to the latent space. The encoder architecture that was selected after the model selection process comprises three graph convolutional layers (ConvGNN) with varying output channels ($15 \rightarrow 32 \rightarrow 64 \rightarrow 48$). Each convolutional layer is followed by a ReLU activation function that introduces non-linearity to the model. Additionally, two dropout layers (with $p=0.3$) were used to regularize the model and prevent overfitting.

The final output of the encoder is then passed to two ConvGNN layers that reduce the dimensionality of the encodings ($48 \rightarrow 24$). These layers are defined as $\mu$ and $\sigma$ (cf., \Cref{eq:vgae_inference}), which are used to create the probability distribution for the latent space. This allows the model to generate new graphs by sampling from this distribution. The decoder component is a standard inner product decoder~\cite{kipf2016variational} which is used to reconstruct the input graph. Notably, the total number of parameters for the entire model is $8,128$, which is relatively small, making the model suitable for edge devices.

\subsection{Graph Encodings Distance}\label{graph_dist}
Since different executions may involve a different number of steps (i.e., nodes), the embeddings obtained from \ourname's model may likely have varying sizes (i.e., $emb_{\text{benign}} \in \mathbb{R}^{n\times L}$ and $emb_{\text{attest}} \in \mathbb{R}^{m\times L}$ , where $n \neq m \vee n = m$, and $L$ is the number of extracted features). As such, it is necessary to employ a metric that can facilitate this comparison.

For this purpose, we have selected the Directed Hausdorff distance (\DHdist)~\cite{hausdorff1914grundzuge}.

\begin{theo}\label{Hausdorff}
    \DHdist captures the distance between two point sets, $A$ and $B$, by taking the maximum of the distances between each point $x \in A$ to its nearest neighbor $y \in B$:
    $$\widecheck{H}(A, B) = \max_{x\in A} \left( \min_{y \in B} \parallel x, y \parallel \right).$$
    where $\parallel \cdot, \cdot \parallel$ represents any norm (e.g., euclidean distance).
\end{theo}

As from Definition~\ref{Hausdorff}, \DHdist can handle sets with different cardinalities. Other approaches, such as averaging pure L2-norms (i.e., euclidean distances) are not applicable in our scenario. Averaging would result in a misleading distance metric, as the result would be impacted by most nodes having a close-to-zero distance to the reference benign trace. This would lead to an increased number of false-negatives.

\section{Implementation}\label{sec:implementation}

In the following sections we delve into the specific techniques and methods used to carry out the key steps of \ourname, emphasizing the significance of each in the model's performance.

\subsection{Tracing and Data Collection}\label{sec:impl:data_collection}
In the following, we describe the methodology employed by our industry partner for data collection in embedded devices. 

We Use DynamoRIO~\cite{dynamorio:phd_thesis} to dynamically instrument prover programs.
Precisely, our tracing prototype instruments a key-exchange application executing on a BeagleBone Black\footnote{\url{https://beagleboard.org/black}\\} with an AM335x 1GHz ARM Cortex-A8 processor. Further, we consider a popular cryptographic library  executing on a Xilinx Ultrascale+ ZCU102\footnote{\url{www.xilinx.com/products/boards-and-kits/ek-u1-zcu102-g.html}} with an ARM Cortex-A53 processor. The CPUs have a 32-bit and 64-bit instruction set architecture, respectively. 
The Embench software and the 40 RIPE executions are traced on an NVIDIA Jetson TX2 NX\footnote{\url{https://developer.nvidia.com/embedded/jetson-tx2-nx}} equipped with a Dual-core NVIDIA Denver 2 64-bit CPU and a quad-core ARM Cortex-A57 CPU.

The instrumentation captures the execution path of the programs in basic-block (BBL) granularity. The execution path is a single path of the respective Complete Control-Flow Graph (CCFG) and represents a single execution of an application. Moreover, as our prototype performs dynamic instrumentation on the binaries, it traces the applications and all the loaded libraries. Thus, it does not introduce interoperability issues with legacy non-instrumented libraries, nor it requires the source code to be available upfront. This approach is applicable to any firmware and software which compiles to binary. Finally, each visited BBL address is stored in a trust anchor isolated to be protected from adversarial tampering. The executed BBLs' addresses are transmitted securely to the verifier. \ourname performs dynamic binary translation to trace control flows of processes in the normal world and stores them in the secure world. To achieve a representative set of execution flows (i.e., a high code path coverage) we traced the applications using a mutation-like fuzzing paradigm. Specifically, we start from a corpus of hand-crafted values and introduce random bit flips to change the input while observing traces that result in new execution paths. After the initial tracing for model training, the same tracing prototype is used to capture the execution path of programs invocation under potential adversary control. Using the same tracing approach for training and inference, we reduce dependencies and data manipulation requirements that improve the verification throughput.

\subsection{Datasets}\label{sec:dataset}
We evaluate \ourname on (i) a popular benchmark suite for embedded devices, Embench~\cite{bennett2022embench}, (ii) the embedded cryptographic library OpenSSL as well as Diffie-Hellman key exchange, (iii) a set of 40 real-world attacks on embedded software, and (iv) a set of synthetically generated control-flow attacks for the first two datasets. All applications are implemented in C/C++. The data collection methodology is described in~\Cref{sec:impl:data_collection} and the extracted dataset statistics are reported in Table~\ref{tab:dataset_stat}. We traced the applications for a minimum of 99 times, with different arguments (cf., \Cref{sec:impl:data_collection}) to acquire a representative set of execution paths, which are used in turn to train and evaluate \ourname. We will evaluate \ourname on a combined dataset size of around $1TB$.

\subsubsection{Embedded Platforms Dataset}
To assess the applicability of \ourname in the context of resource-constrained devices, we build a dataset by tracing the benchmarks in Embench~\cite{bennett2022embench}\footnote{\url{https://github.com/embench/embench-iot}}, a representative suite of 18 benchmarks for embedded devices. The benchmarks are meant for embedded devices, the length of the traces varies between $112\text{k}$ steps up to $657\text{k}$ steps. The traced benchmark counts approximately $2\text{k}$ nodes and $2.4\text{k}$ edges.

\subsubsection{Cryptographic Dataset}
We also evaluate mature encryption algorithms implemented in the OpenSSL v1.1.1f\footnote{\url{https://www.openssl.org/news/openssl-1.1.1-notes.html}} library which are typically used by embedded devices for maintaining data confidentiality. Specifically, we implemented a wrapper application in C/C++ that encrypts 1KB block with 56-bit DES, 64-bit DES-X, 64-bit GOST and 128-bit AES using randomly generated keys. 
The tracing of these applications resulted in trace lengths of $13\text{M}-16\text{M}$ execution steps with approximately $9\text{k}$ edges and $7\text{k}$ nodes.
Additionally, we include the popular Diffie–Hellman key exchange algorithm\footnote{\url{https://github.com/thejinchao/dhexchange}} in the analysis. %The tracing is done on the BeagleBone Black device. 
While this implementation is relatively simple, its trace length is approximately $223\text{k}$ execution steps long with around $2\text{k}$ edges and nodes. 
Note that, prior work on CFA for embedded devices, e.g., C-Flat~\cite{abera2016c} and BLAST~\cite{yadav2023whole} evaluated on the Open Syringe Pump library with just $332$ edges. Thus, we consider our datasets to be much more complex.

\begin{table}[ht]
\centering
\caption{For each dataset, the numbers are reported as \res{mean}{std}. \# Exe. indicates the number of traced executions. Trace Len. indicates the length of the traced executions (i.e., amount of execution steps).}

\resizebox{.82\columnwidth}{!}{
\begin{tabular}{l|ccccc}

\textbf{Dataset} & \textbf{\# Exe.} & \textbf{Trace Len.} &\textbf{\# Nodes} & \textbf{\# Edges} & \textbf{Degree}  \\ \hline

Diffie-Hellman & 1000 &\res{2.23 \times 10^5}{6.01 \times 10^3} &\res{1921}{7} & \res{2295}{9} & \res{2}{0}\\
DES & 260 & \res{1.44 \times 10^7}{2.26 \times 10^3} &\res{7194}{3} &\res{9361}{3} & \res{3}{0} \\ 
DESX & 260 & \res{1.47 \times 10^7}{3.00\times 10^3} &\res{7386}{2} & \res{9677}{3}  &  \res{3}{0}  \\ 
GOST & 260 & \res{1.36\times 10^6}{2.35\times 10^3} &\res{7211}{52} & \res{9301}{71}  &  \res{3}{0}  \\ 
AES & 260 & \res{1.62\times 10^6}{1.50\times 10^3}&\res{7140}{7} & \res{9217}{9}  &   \res{3}{0} \\
aha-mont64 & 99 & \res{3.98\times 10^5}{41} & \res{1539}{12} & \res{1783}{12} & \res{2.3}{0} \\ 
statemate & 99 & \res{1.45\times 10^5}{2.2\times 10^2} & \res{1575}{25} & \res{1822}{26} & \res{2.3}{0} \\ 
primecount & 99 & \res{1.78\times 10^6}{3.1\times 10^2} & \res{1533}{19} & \res{1786}{21} & \res{2.3}{0} \\ 
sglib-combined & 99 & \res{6.57\times 10^5}{36} & \res{1763}{1.6} & \res{2110}{1.8} & \res{2.4}{0} \\ 
minver & 99 & \res{1.12\times 10^5}{82} & \res{1565}{1.7} & \res{1831}{1.7} & \res{2.3}{0} \\
edn & 99 & \res{3.84\times 10^5}{18} & \res{1586}{1.7} & \res{1848}{1.8} & \res{2.3}{0} \\ 
crc32 & 99 & \res{5.4\times 10^5}{3.4\times 10^4} & \res{1516}{16} & \res{1760}{17} & \res{2.3}{0} \\ 
matmult-int & 99 & \res{4.2\times 10^5}{3.4} & \res{1561}{1.6} & \res{1819}{1.9} & \res{2.3}{0} \\
picojpeg & 99 & \res{4.4\times 10^5}{33} & \res{1807}{1.6} & \res{2148}{1.8} & \res{2.4}{0} \\ 
md5sum & 99 & \res{3.42\times 10^5}{3.1\times 10^3} & \res{2002}{2.0} & \res{2319}{2.6} & \res{2.3}{0} \\ 
nsichneu & 99 & \res{7.87\times 10^5}{62} & \res{2152}{1.7} & \res{2399}{1.9} & \res{2.2}{0} \\ 
wikisort & 99 & \res{2.27\times 10^5}{1.1\times 10^2} & \res{1680}{1.6} & \res{1996}{2.1} & \res{2.4}{0} \\ 
tarfind & 99 & \res{1.7\times 10^5}{95} & \res{1544}{20} & \res{1791}{21} & \res{2.3}{0} \\ 
huffbench & 99 & \res{5.13\times 10^5}{2.9\times 10^4} & \res{1641}{17} & \res{1929}{19} & \res{2.4}{0} \\ 
cubic & 99 & \res{1.6\times 10^5}{15} & \res{1881}{1.7} & \res{2232}{2.4} & \res{2.4}{0} \\ 
nettle-aes & 99 & \res{8.95\times 10^4}{40} & \res{1560}{1.7} & \res{1814}{1.9} & \res{2.3}{0} \\ 
slre & 99 & \res{6.37\times 10^5}{72} & \res{1641}{25} & \res{1922}{27} & \res{2.3}{0} \\ 
qrduino & 99 & \res{4.73\times 10^5}{3.9\times 10^4} & \res{1898}{23} & \res{2301}{25} & \res{2.4}{0} \\ 
\end{tabular}
}

\label{tab:dataset_stat}
\end{table}

\subsubsection{Real-World Attack Traces}\label{subsubsec:real_attacks}
To evaluate the feasibility of our approach, we manually added a ROP backdoor to the key exchange software. 
Specifically, we introduce a buffer overflow vulnerability on a stack pointer that, when exploited, enables an adversary to modify the function return address to a chosen location. This technique is similar to existing ROP vulnerabilities. We then generate attack traces by exploiting this vulnerability while tracing the program. 

Additionally we applied the RIPE framework exploit collection~\cite{wilander.ripe}, which our industry partner extended to ARM platforms, to the Embedded Platform Software. The RIPE benchmark consists of a vulnerable program and a set of 850 buffer overflow exploits with various techniques.

The framework is configurable in terms of attack type (i.e., ROP and ret2libc), overflow type (i.e., direct, indirect), attack location (i.e., stack or heap), target code pointer (e.g., function pointer, return address) and function exploited (e.g., strcpy, strcat). To verify whether our approach is affected by different ROP chain lengths we further extended RIPE by a gadget length configuration option.

To collect traces from different software than the one provided by RIPE, we extended the framework to include benign application logic. We use the picojpeg benchmark from the Embench suite as a representative benign application. We therefore include a configuration option to disable the exploitation of the vulnerability and continue with executing the benign application. 

As a representative set of attacks we chose 40 different combinations of configuration options (cf., \Cref{tab:ripe_details}). As \ourname does not discern between different code pointers or memory locations but only considers the difference in the traced execution path, we only evaluate all attacks containing different buffer overflow techniques.
Thus, the attack traces reflect real-world changes in the execution path addresses, which in turn can be used to evaluate the accuracy of \ourname.

\subsubsection{Synthetic Trace Generation}\label{subsubsec:rop_dop_gen}
As discussed earlier, acquiring traces while the program is under attack needs vulnerabilities to be introduced or the exploitation of existing (known) ones.
However, this is both time-consuming and limits attack traces to existing vulnerabilities and known Control-Flow hijacking techniques.
Instead, we observed that the attack traces and benign traces differ only in the nodes related to the ROP or ret2libc attack. While the buffer overflow implementation of the injected vulnerability is specific to this type of software, the attack is software agnostic, i.e., it corrupts a code pointer that causes a change in the execution integrity and is reflected in the traces.
Thus, we propose a novel synthetic attack trace generation approach, which is based on traces extracted from the benign execution and modified to accurately simulate ROP and DOP attacks in the most generic way.
In the following, the algorithms are presented which are based on the earlier acquired data for ROP attacks and the description by Hu~\etal~\cite{hu2016data} for the DOP attack traces.

To establish a reliable baseline and evaluation, traces showing malicious behavior in a known region in the trace are needed. Hence, we inject traces of a ROP or DOP attack into known benign traces. As described in \Cref{sec:dataset}, we studied the effects of these attacks on a CFG (cf., \Cref{background:cfg}) and implemented two algorithms to mimic the effects of ROP and DOP attacks. In addition, we parameterized the algorithms to be able to generate outputs of different lengths, positions and repetitions.

\begin{algorithm}
\caption{ROP Generator}\label{alg:ROP}
\begin{algorithmic}
\Require $pos$ \Comment{position in trace to add ROP traces}
\Require $inserts$ \Comment{amount of malicious steps to insert}
\Require $trace$ \Comment{list of steps as addresses}
\Ensure $trace$ \Comment{list with added ROP}
\State $n \gets pos$
\While{$n \neq pos + inserts$}
    \State $elem \overset{{\scriptscriptstyle\$}}{\gets} trace$ \Comment{sample uniformly from trace}
    \State $trace \gets trace_0, ..., trace_{n-1}, elem, trace_{n}, ...,\linebreak trace_{|trace|}$
    \State $n \gets n + 1$
\EndWhile
\State \Return $trace$
\end{algorithmic}
\end{algorithm}

\Cref{alg:ROP} shows the pseudo-code of the algorithm used to generate traces of an execution of software under ROP attack. The function takes as arguments the list of execution steps (i.e., addresses) \emph{trace}, the number of steps \emph{inserts} and the position \emph{pos} in the trace to insert. The algorithm uniformly samples \emph{inserts} elements of the \emph{trace} list and adds it to the \emph{trace} list at position \emph{pos}. With this, the \emph{trace} list contains a block of re-used addresses with (probably) new jumps (between addresses) which did not exist before in the input \emph{trace}. As we don't know which address belongs to which processor instruction, uniformly sampling addresses is the most generic way of creating ROP traces. Usually, ROP gadgets should end with a return statement (or e.g., jump, branch), making our generated traces even more challenging to detect, as they lack structure.

\begin{algorithm}
\caption{DOP Generator}\label{alg:DOP}
\begin{algorithmic}
\Require $pos$ \Comment{position in trace to add DOP traces}
\Require $inserts$ \Comment{amount of malicious steps to insert}
\Require $repeats$ \Comment{repetitions of inserted steps}
\Require $trace$ \Comment{list of steps as addresses}
\Ensure $trace$ \Comment{list with added DOP}

\State $atk \gets \{atk_0, ..., atk_{inserts} | \;atk_i \in [trace_0, trace_{pos}] \subseteq trace,\; 
    \forall\; atk_i, atk_{i+1}: atk_i = trace_j \wedge atk_{i+1} = trace_{j+1}, \;\exists\; trace_{pos-1}, atk_0: trace_{pos-1} = trace_j \wedge atk_0 = trace_{j+1}, \;\exists\; trace_{pos}, atk_{inserts}: trace_{pos} = trace_j \wedge atk_{inserts} = trace_{j+1} \}$
\State $trace \gets trace_0, ..., trace_{pos-1}, atk_0, ..., atk_{inserts},\linebreak trace_{pos}, ..., trace_{|trace|}$

\State $n \gets repeats$
\While{$n \neq 0$}
    \State $trace \gets trace_0, ..., trace_{pos-1}, atk_0, ..., atk_{inserts},\linebreak\hspace*{\algorithmicindent} trace_{pos}, ..., trace_{|trace|}$
    \State $n \gets n - 1$
\EndWhile
\State \Return $trace$
\end{algorithmic}
\end{algorithm}

\Cref{alg:DOP} shows the pseudo-code of the algorithm used to generate traces of an execution of software under DOP attack. It has the same inputs as \Cref{alg:ROP}, including an additional \emph{repeats} argument, which controls how often the block of DOP steps is repeated. This is used to mimic the behavior of DOP attacks on cryptographic software to, e.g., locate a secret key in memory~\cite{hu2016data}. The algorithm tries to find a sequence of steps (\emph{atk}) of length \emph{inserts} from the already visited nodes (until position \emph{pos} in \emph{trace}) where each pair of steps in the sequence must also be a pair in the \emph{trace} list (i.e., all jumps must also occur in the benign trace). In addition, the first and last element of the sequence must fit into the benign trace at position \emph{pos} (i.e., also these jumps occur in the benign trace). If no solution is found, the algorithm is re-run with a different \emph{pos} argument (not depicted in the pseudocode). Finally, the DOP traces are repeatedly added \emph{repeats} times to the \emph{trace} list on position \emph{pos}. This approach creates DOP traces by adding a new path to the \emph{trace} list, which only utilizes jumps already taken by the benign execution (i.e., not creating new edges), therefore not changing the Control-Flow Graph (CFG) but changing the control flow by executing jumps in a sequence never done before by the benign program.

\subsection{Data Preprocessing}\label{subsec:data_pre}

The objective of the data preprocessing phase is to prepare the raw data obtained from the tracer for subsequent machine learning steps. A trace consists of a sequence of memory addresses that the program counter has traversed during execution. The first step in the preprocessing phase is to convert the list of execution steps, denoted as $\text{Exe}_\text{steps}=\{s_1, s_2, ..., s_n\}$, into a graph representation. Specifically, we construct an edges list~$E = \{e_1, e_2,\, ...,\, e_n\}$, where $e_i=(v_j, v_k)$ represents an edge between two nodes, $v_j$ and $v_k$ belonging to $V$, in the graph. Each node in the graph represents a unique memory address referenced in the execution steps.

Once the edges list $E$ is constructed, it is further processed to convert it into graph connectivity in coordinate format (COO), referred to as $A$. This is achieved by applying the transpose operation $A = E^T$. This conversion is done as the COO format is more suitable for the subsequent machine-learning step. Furthermore, while we are iterating on $\text{Exe}_\text{steps}$ and building the edges list, we also conduct the feature extraction phase in parallel. This is an essential step as it allows the addition of extra information to the graph representation that can be used to improve the model's performance.

\subsubsection{Feature Extraction}\label{feature_extr}

In the feature extraction phase, we aim to identify various characteristics of each node $v_i \in V$ in the graph $G$, which represents the control flow of one software execution. Firstly, we construct the Partial Control-Flow Graph (PCFG) of this single execution by generating an edge list, represented as an array $E \in \mathbb{N}^{n\times 2}$, where $n$ is the number of executed unique addresses (i.e., vertices) for a given software.

One such characteristic is the vertex degree, denoted as $\text{deg}(v_i)$, which is the number of edges that are incident to $v_i$. This feature provides information about the number of edges that are connected to a given node, which can indicate the centrality of a node in the graph. Another feature is the number of visits, which is the number of times the corresponding address is accessed during the execution. This feature can provide information about the overall activity of a node in the execution and its frequency of visits.

We also identify the first and last visit of each node, considering the execution steps. This feature can provide information about the lifetime of a node in the execution. In addition, we determine the number of incoming and outgoing edges, which is the number of edges that end in $v_i$ and the number of edges that start from $v_i$, respectively. 

We also compute the frequency of visits, which is the number of visits over the total number of execution steps, as well as the time of use, which is the difference between the last visit and the first visit, i.e., how long that node has been used during the execution. These features provide information about the relative importance of a node in the execution and the amount of time spent on it. 

Furthermore, we calculate the standard deviation of all the visit time to a node, given the list of visits of the node $i$, $\text{vis}_i=\{s_i, s_j, ..., s_k\}$, where $s_k$ is the k-th step where the address represented by $v_i$ is seen for the last time and the mean of the distance between visits, given the list of visits of the node $i$, $\text{vis}_i=\{s_i, s_j, ..., s_k\}$, we compute the mean of the distance between each couple of visits, i.e.,~$\text{mean}\left((s_j-s_i), ..., (s_k-s_j)\right)$. These features provide information about the temporal relations between visits, indicating the distribution of the visits.

We also determine the mean number of visits of incoming and outgoing neighbors for all the nodes that have an edge ending in $v_i$ and for all the nodes that are the ending of the edges starting from $v_i$, respectively. These features provide information about the activity of the neighboring nodes, indicating the level of activity of the surrounding nodes. 

After extracting all the features, we normalize them according to the trace length to ensure that the values are comparable across different execution traces. Furthermore, we categorize these features according to their intended use in the scope of our study and present them in Appendix in \Cref{tab:features_dop_rop}. It is crucial to note that all the extracted features are computationally efficient: we estimated that the algorithmic complexity for their extraction can be approximated to $O(n)$, where $n$ is the length of the trace. This enables fast data preprocessing and efficient run-time attestation, which aligns with the goal of our study.

\subsection{Model Training}\label{subsec:implementation_model_tr}

In the training process, our \ourname's conducts an unsupervised model training utilizing a single arbitrary execution. The objective of this training is to achieve generalization to the benign traces, so that the distance between benign traces (i.e., $\text{trace}_i$) and the training trace (i.e., \tracezero) has to be  as close as possible to 0, i.e., $d(\text{trace}_0, \text{trace}_i)\rightarrow 0$. To accomplish this, the VGAE model is trained for a maximum of $3,000$ epochs, incorporating an early stop mechanism that monitored the Average Precision (AP) and Area-Under-Curve (AUC) values, with a patience of $500$ epochs. As outlined in \Cref{subsec:background_vgaes}, the goal is to optimize the variational lower bound, as described in \Cref{eq:variational_low_bound}). 

Additionally, we carried out an extensive hyperparameter selection process to find the optimal set that resulted in the best performance of \ourname. Specifically, we tested the system on multiple hyperparameter combinations, e.g., different sizes of the latent space, different number of layers in the encoder, and the number of channels in each layer.
To regularize the learning process, a decaying learning rate (\textit{lr}) scheme is adopted, starting with $lr=0.01$ and decreasing by a factor of $3$ every $150$ epochs until epoch $750$, after which it is fixed. Further, we leverage dropout layers in the encoder (cf., \Cref{fig:VGAE_encoder}b). This way, we effectively regularize the learning process and prevent it from getting stuck in a local minimum, thereby allowing for continuous improvement. The use of these regularization techniques, in conjunction with the optimal set of hyperparameters, resulted in a well-performing model that is able to generalize to unseen benign traces.

It is especially important to note that \ourname's model settings are fixed, i.e., the determined optimal settings are therefore used for the evaluation of all datasets: \ourname aims to work ``out of the box'' without requiring a model selection phase when changing datasets. It only needs a quick unsupervised training phase on one benign execution trace.

\subsection{Attestation Threshold}\label{subsec:implementation_threshold}
A reliable attestation requires a carefully designed attestation threshold, for determining the integrity of an execution.
To accomplish this, we evaluate the number of traces needed for defining a reliable threshold by carrying out a systematic ablation study which we report in \Cref{app:threshold_ablation_study}. We observe that the selection of just $10$ random traces maximizes the performance of \ourname while maintaining a low false-positive-rate.
For each of these traces, the distance to \tracezero is calculated and the threshold $t$ is established using the following formula:

\begin{equation}
    t_\text{attestation} = \text{mean}(d_\text{ValSet}) + 2\times \text{std}(d_\text{ValSet})
\end{equation}

where $d$ is the Directed Hausdorff distance (cf., Definition~\ref{Hausdorff}), $d_\text{ValSet}$ represents the list of distances between the graphs in the validation set and \tracezero. It is important to note that while a handful of traces from the validation set is used for thresholding, only \emph{one} trace is utilized for training. 

During the attestation phase, the embeddings of the execution to be verified are obtained through the encoder. The distance $d$ between these embeddings and the reference execution, represented by \tracezero, is subsequently calculated. If this measure exceeds a predefined threshold, the execution is deemed to be malicious.

\section{Evaluation}\label{sec:evaluation}

In the following sections, we evaluate the effectiveness of \ourname in detecting different attacks, aiming to answer the research questions we formulated in \Cref{sec:intro}.

\subsection{Evaluation Metrics}
 To evaluate the effectiveness of \ourname, we utilize a variety of metrics, including Precision, Recall, and F1-score. Additionally, we use the False Positive Rate (FPR) to provide more insight into the detailed performance of the model. The choice of these metrics is driven by the need for a precise assessment of the performance of the attestation system: we particularly care about knowing how reliable the system is.

\subsection{Results}\label{subsec:eval_results}
To address the above research questions, we test \ourname against traces containing ROP and DOP attack executions for each dataset.
For the following results, the experiments are repeated 100 times, and the attestation threshold was computed based on the distance to \tracezero of 10 random benign traces as described in \Cref{subsec:implementation_threshold}.

\subsubsection{ROP}

As described in \Cref{subsubsec:real_attacks} we utilized the RIPE framework to collect 440 traces in total, including 40 different real-world ROP and ret2libc attack traces (cf., \Cref{tab:ripe_details} in Appendix). 

The results in \Cref{fig:RIPE} in the Appendix clearly depict \ourname's capability of correctly separating benign and attack traces in every case. Formally, this means that the embeddings of the real-world attacks under test, are diverging significantly from the reference embeddings. This leads us to conclude that the real-world attack traces generated by adding RIPE exploits are not executed stealthily (i.e., resulting in a shell, terminating execution). 

Instead, as described in \Cref{subsubsec:rop_dop_gen}, our attack generator's traces are much more sophisticated as the simulated ROP attack returns the execution flow to the benign program (e.g., instead of terminating the execution), requiring more advanced detection capabilities. 

\begin{insight}{Insight 1}\label{insight1}
\textbf{RQ2:} The embeddings of the graph preserve the characteristics of an execution trace. Specifically, during the embedding process: $$f: v_i \rightarrow z_i \in \mathbb{R}^L$$ 
\ourname's model is acting as embedding function $f$, mapping the execution graph's node $v_i$ to embeddings $z_i$ keeping a close correspondence between high and low dimensional data~\cite{xu2021understanding}. Thanks to this property the link between the state of the execution and the latent representation is preserved.
\end{insight}

We, therefore, generate multiple ROP attacks with different attack chain lengths. In particular, we generate 50 malicious traces by adding ROP-type steps for each of the following lengths to a random benign execution that is recognized by \ourname as benign: $5$, $10$, $15$, $30$, $40$, $50$, $75$, $100$, $150$, $200$, $250$, $350$ and $500$. 

As shown in \Cref{tab:rop_eval}, in the case of Diffie-Hellman, our key exchange dataset (cf., \Cref{sec:dataset}), we can achieve an F1-Score of 92.58\% and a False-Positive-Rate (FPR) of 7.12\%.
The performance for OpenSSL is better, achieving an F1-Score of at least 97.95\% 
For Embench, the results generally show better performances than the cryptographic dataset, as the binaries are less complex. In the case of ``sglib-combined'' and ``minver'', we can achieve even perfect detection and null false-positive. On average, we achieve an F1-Score 98.03\% and an FPR of 3.19\%.

We show that \ourname is better at detecting ROP attacks rather than to detect DOP attacks, as they modify the Control-Flow Graph's (CFG) structure (i.e., creating new edges), leading to better performance.
In this case, the extracted features help to highlight the ongoing attack. Still, the graph plays a significant role (i.e., the adjacency matrix will change after the attack). In \Cref{fig:rop_eval}, these results are graphically represented.

\subsubsection{DOP}
As described in \Cref{subsubsec:rop_dop_gen}, we generate 100 DOP attacks with a fixed length of around 2000 execution steps. 
As reported in \Cref{tab:dop_eval}, in the case of Diffie-Hellman (cf., \Cref{sec:dataset}), we can achieve an F1-Score of 69.22\% and a False-Positive-Rate (FPR) of 7.12\%.
For the OpenSSL dataset, achieving an F1-Score of at least 86.41\% for all the ciphers is possible. Among these algorithms, DES and DESX deliver the best results with an F1-Score of 90.10\% and 89.81\%, respectively.
For Embench, the results show better performances than the cryptographic dataset, as the binaries are less complex. Regarding ``sglib-combined'', we can achieve perfect detection and 0 false-positive. On average, we achieve an F1-Score 91.01\% and an FPR of 3.19\%.

In contrast to ROP attacks, the changes made by the attack to the CFG are much less evident (i.e., DOP attacks do not create new edges; they execute a new path with benign steps) and, therefore, are much harder to detect. Moreover, for detecting DOP attacks, the feature extraction phase plays a fundamental role, as a DOP attack can only be recognized from the nodes' features (e.g., the number of visits) and not from the graph's structure itself (i.e., the adjacency matrix, which stays unchanged). 
In \Cref{fig:dop_eval}, the results are graphically represented.

\begin{insight}{Insight 2}
\textbf{RQ3:} The embedding changes are sufficient even to detect the most stealthy changes to the execution trace made by ROP and even DOP attacks. Formally, the model effectively produces embeddings such that, the separation between benign embeddings \footnotesize{$d_b=d(Emb_\text{ref}, Emb_\text{benign})$} \normalsize and malicious embeddings \footnotesize{$d_a=d(Emb_\text{ref}, Emb_\text{attacks})$} \normalsize is \footnotesize{$d_a-d_b \gg t$}, \normalsize where $t$ is the attestation threshold, is enough to result in accurate detection.
\end{insight}

\begin{table}[ht]
\centering
\caption{For each dataset, the following evaluation metrics are reported: false-positive-rate (FPR), precision (Pr), recall (Re), and F1-Score. All the values are in percentages. The values are a mean of 100 experiments.}

\resizebox{0.87\columnwidth}{!}{
\begin{tabular}{l|c|ccc|ccc}
                   &              & \multicolumn{3}{c|}{\textbf{ROP}}          & \multicolumn{3}{c}{\textbf{DOP}}       \\
\textbf{Dataset}   & \textbf{FPR} & \textbf{Re.} & \textbf{Pr.} & \textbf{F1} & \textbf{Re.} & \textbf{Pr.} & \textbf{F1} \\
\hline
DH                   & 7.12         & 95.44        & 90.59        & 92.58       & 86.03     & 62.51        & 69.22       \\
AES            & 0.78         & 99.06        & 99.69        & 99.37       & 78.92     & 95.88        & 86.41       \\
DES            & 9.10         & 99.70        & 96.58        & 98.09       & 100.0     & 82.79        & 90.10       \\
DESX           & 8.82         & 99.30        & 96.66        & 97.95       & 98.89     & 82.98        & 89.81       \\
GOST           & 1.51         & 99.57        & 99.40        & 99.48       & 82.81     & 93.49        & 86.60       \\
aha-mont64     & 0.19         & 100.0        & 99.97        & 99.99       & 100.0     & 98.50        & 99.18       \\
crc32          & 7.23         & 96.28        & 98.93        & 96.51       & 85.00     & 70.07        & 75.94       \\
cubic          & 6.56         & 100.00       & 99.05        & 99.52       & 99.72     & 94.52        & 96.93       \\
edn            & 3.25         & 99.72        & 99.52        & 99.62       & 99.27     & 91.89        & 95.29       \\
huffbench      & 6.48         & 99.48        & 99.06        & 99.26       & 94.12     & 94.06        & 92.01       \\
matmult-int    & 0.28         & 97.02        & 99.96        & 98.47       & 96.72     & 99.65        & 98.12       \\
md5sum         & 3.84         & 99.99        & 99.44        & 99.71       & 99.47     & 96.44        & 97.92       \\
minver         & 0.00         & 100.0        & 100.0        & 100.0       & 99.97     & 100.0        & 99.98       \\
nettle-aes     & 2.96         & 100.0        & 99.57        & 99.78       & 99.51     & 97.23        & 98.35       \\
nsichneu       & 2.90         & 89.90        & 99.53        & 94.43       & 100.0     & 68.90        & 81.41       \\
picojpeg       & 0.08         & 100.0        & 99.99        & 99.99       & 100.0     & 99.92        & 99.96       \\
primecount     & 3.69         & 72.00        & 99.26        & 83.46       & 53.00     & 94.19        & 67.76       \\
qrduino        & 2.57         & 92.42        & 99.6         & 95.16       & 78.00     & 75.94        & 76.95       \\
sglib-combined & 0.00         & 100.0        & 100.0        & 100.0       & 100.0     & 100.0        & 100.0       \\
slre           & 8.06         & 100.0        & 98.84        & 99.41       & 100.0     & 93.23        & 96.40       \\
statemate      & 1.68         & 100.0        & 99.75        & 99.87       & 100.0     & 98.44        & 99.21       \\
tarfind        & 1.08         & 100.0        & 99.84        & 99.92       & 98.57     & 98.97        & 98.75       \\
wikisort       & 6.50         & 99.97        & 99.05        & 99.51       & 99.83     & 94.29        & 96.93      
\end{tabular}
}
\label{tab:dop_eval}
\label{tab:rop_eval}
\end{table}

\subsection{Runtime Overhead}\label{subsec:eval_runtime}

We evaluate \ourname's runtime overhead of the operations carried out by the verifier in terms of storage and computation time for preprocessing, training, and inference phases. In \Cref{sec:eval_runtime}, we report the results.

\section{Security Considerations}\label{sec:sec}
In this section, we analyze two main security aspects: (i) implications of not assuming the completeness of the control-flow graph and (ii) considerations on our performance detecting code reuse attacks.

\subsection{CFG Completeness}\label{sec:sec_completeness}
In \Cref{background:cfg}, we analyzed the impracticability of assuming the completeness of a CFG.

On these premises, we stress that the related works such as 
C-FLAT~\cite{abera2016c}, Atrium~\cite{zeitouni2017atrium}, Lo-fat~\cite{dessouky2017fat}, Litehax~\cite{dessouky2018litehax}, OAT~\cite{sun2020oat}, Recfa~\cite{zhang2021recfa}, Tiny-CFA~\cite{nunes2021tiny}, and BLAST~\cite{yadav2023whole} cannot be considered practical for control-flow attestation, as they rely on the availability of a CCFG. Specifically, this implies that these approaches (i) exhibit a high false-positive rate proportional to the undiscovered CFG regions and (ii) miss all attacks that might occur in these regions.

Unlike these approaches, we do not rely on assumptions about the PCFG. \ourname addresses this challenge by being independent of the completeness of the PCFG. Specifically, \ourname operates on execution graphs (i.e., using only \emph{one} execution trace) by exploiting a deep unsupervised geometric model, which learns a program's benign behavior and patterns. 
This is proven by, Insight 1 where we showed that \ourname's model acts as an embedding function $f$ which maps high-dimensional data $V$ (i.e., the execution traces) to a low-dimensional space $Z$, extracting latent features that capture the execution's behaviour while, ultimately, keeping a close link between trace and embeddings $f: V \to Z$.

Furthermore, in terms of performance, \ourname achieved a minimal false-positive rate, as low as 0.39\% in the best scenario. This result significantly surpasses the related works when considering the above mentioned factors, making it suitable for real-world applications.

Moreover, \ourname only requires the basic block addresses traversed by the program counter for attestation. Thus preventing information leakage to the verifier, as this does not disclose any information about the prover’s computation content to the verifier~\cite{molnar2006program}.

\subsection{CRA Detection}
Our empirical evaluation has shown that \ourname is effective in detecting Code Reuse Attacks (CRA) with a high degree of accuracy (cf., \Cref{subsec:eval_results}). 
In addition to the empirical evidence, this section conceptually discusses the effectiveness of our approach in detecting (i) control-data attacks and (ii) non-control data attacks.

To perform control-data attacks, such as ROP attacks, an attacker must alter the control flow of a program, creating new transitions (cf., \Cref{subsec:background_memory}). This results in illegal edges in the CFG. \ourname exploits the graph representation of an execution trace for extracting low dimensional embeddings that reveal the execution behavior, leveraging the concepts behind Definition \ref{def:embs} through probability sampling, as described in \Cref{eq:vgae_inference}. Thus, when comparing a malicious execution with a benign one, \ourname expresses the (dis)similarity (i.e., distance) of these two executions. High distance values imply a significant alteration of the CFG.

\begin{insight}{Insight 3}
\textbf{RQ4:} Structural (Behavioural) changes in the graph are preserved by the embeddings (\textbf{RQ2}, Definition \ref{def:embs}) and are related by \ourname to an execution being attacked by a ROP (DOP) attack.
\end{insight}

While performing non-control data attacks, such as DOP attacks, an attacker would reuse benign transitions to manipulate the execution flow. This results in a more stealthy attack, as the execution graph will only consist of benign edges. While converting an execution trace into an execution graph, \ourname extracts numerous descriptive features (cf., \Cref{feature_extr}) that reveal underlying properties of the execution flow. 
Through the embedding comparison, \ourname will show a significant distance between the benign execution and the compromised one, resulting in attack detection. For example, a typical pattern in DOP attacks is the repeated execution of specific basic blocks, which will result in an unusual increase in the respective features, e.g., \emph{Frequency of visits} (cf., \Cref{tab:features_dop_rop}).

\begin{insight}{Insight 4}
\textbf{RQ1:} Our approach can reliably detect CRAs (\textbf{RQ3}) while providing sufficient security assumptions, therefore being able to relax the assumption of access to a complete control-flow graph.
\end{insight}

%\vspace{-2.2em}
\section{Related Work}\label{sec:related}

While machine learning has been utilized for attestation, to the best of our knowledge, we are not aware of any Control-Flow Attestation (CFA) scheme that uses Graph Neural Networks (GNN) and exploits the natural correspondence between execution trace, execution graphs and execution embeddings. We are also unaware of approaches able to be trained unsupervised on only \emph{one} known benign trace.

We focus therefore on existing CFA methods, the different works on static and run-time attestation, and the recent ML-based attestation schemes. We report an overview comparison between related works in \Cref{tab:related_work_assumption}. 

\subsection{\textbf{Static Attestation}}

Static Attestation is a process by which the security properties of a system or device are verified based on the analysis of its design and implementation rather than by observing its runtime behavior. This means the system or device is evaluated based on its intended functionality instead of observing its actual behavior at runtime. 
A popular approach for Static Attestation using Software is hashing~\cite{seshadri2004swatt, seshadri2006scuba, chen2017secure, seshadri2005pioneer, li2011viper} or timing of an algorithm working on~\cite{seshadri2008sake} the firmware in flash or memory. Utilizing a hardware root of trust such as a Trusted Platform Module (TPM), it is possible to attest a device running a known benign software~\cite{krauss2007detecting, agrawal2015program}.

Other approaches not only measure the critical software but also protect it at runtime using isolation~\cite{eldefrawy2012smart, koeberl2014trustlite, brasser2015tytan, carpent2018remote}, others leveraging Physically Unclonable Functions (PUF)~\cite{aman2018att}, or extra hardware~\cite{aman2021prom} to attest memory at runtime.

\subsection{\textbf{Control-Flow Attestation (CFA)}}

\begin{table*}[ht]
\centering
\caption{Comparison of assumptions of related work.\\ \checkmark: the property is a requirement, \cross: not a requirement, (\checkmark): not validated and, \NA: not applicable.\\ \textbf{Complete CFG}: The approach requires a complete CFG, \textbf{Custom HW}: custom made hardware, \textbf{Measurements}: a-priori measurements of memory or features, \textbf{Memory Access}: access to the provers program memory, \textbf{Platform Specific}: is only applicable in specific scenarios or platforms, \textbf{Heavy Bin. Modification}: needs heavy binary re-writing, \textbf{Only specific software}: can only attest/protect specific software, or \textbf{Presence of Feature}: depends on specific hardware features.
}

\resizebox{\linewidth}{!}{

\begin{tabular}{c|c|cccccccc|cc}
& & \multicolumn{8}{c}{\textbf{Assumptions}} & \multicolumn{2}{c}{\textbf{Detection Capabilities}} \\
 
&  & \textbf{Complete} & \textbf{Memory} &  & \textbf{Custom} & \textbf{Presence of} & \textbf{Platform} & \textbf{Heavy Bin.} & \textbf{Only Specific} & \textbf{Control-Flow} & \textbf{Data-Only} \\ 

\multicolumn{1}{c|}{\textbf{Category}} & \multicolumn{1}{c|}{\textbf{Approach}} & \textbf{CFG} & \textbf{Access} & \textbf{Measurements} & \textbf{HW} & \textbf{Features} & \textbf{Specific} & \textbf{Modification} & \textbf{SW} & \textbf{Attack} & \textbf{Attack} \\ 

\hline

\multirow{3}{*}{\specialcell{\textbf{Security}\\\textbf{Architecture}}}  &  Smart~\cite{eldefrawy2012smart} &  \NA  &  \NA  &  \NA  &  \checkmark  &  \checkmark  &  \checkmark  &  \NA  &  \checkmark  &  \checkmark  &  \cross \\
 &  Trustlite~\cite{koeberl2014trustlite} &  \NA  &  \checkmark  &  \NA  &  \checkmark  &  \checkmark  &  \checkmark  &  \NA  &  \cross  &  \NA  &  \NA \\
 &  Tytan~\cite{brasser2015tytan} &  \cross  &  \checkmark  &  \checkmark  &  \checkmark  &  \checkmark  &  \checkmark  &  \cross  &  \cross  &  \checkmark  &  \checkmark \\

\hline

\multirow{5}{*}{\specialcell{\textbf{Malware}\\\textbf{Detection}}}  &  Numchecker~\cite{wang2013numchecker}  &  \cross  &  \cross  &  \checkmark  &  \cross  &  \checkmark  &  \checkmark  &  \cross  &  \checkmark  &  \checkmark  &  \cross \\
  
  &  SWARM~\cite{carpent2018remote} &  \cross  &  \checkmark  &  \checkmark  &  \checkmark  &  \checkmark  &  \checkmark  &  \cross  &  \cross  &  \cross  &  \cross \\
  
  &  Att-auth~\cite{aman2018att}  &  \cross  &  \checkmark  &  \checkmark  &  \checkmark  &  \checkmark  &  \checkmark  &  \cross  &  \cross  &  \cross  &  \cross\\
  
  &  Prom~\cite{aman2021prom}  &  \cross  &  \checkmark  &  \checkmark  &  \checkmark  &  \cross  &  \checkmark  &  \cross  &  \cross  &  \cross  &  \cross \\

\hline

\multirow{1}{*}{\textbf{CFI}}  &  Cfimon~\cite{xia2012cfimon}  &  \cross  &  \cross  &  \checkmark  &  \cross  &  \checkmark  &  \checkmark  &  \cross  &  \cross  &  \checkmark  &  \cross \\% HPC

\hline  
\multirow{3}{*}{\textbf{ML}}  &  Hu \etal~\cite{hu2019probability}  &  \checkmark  &  \cross  &  \checkmark  &  \cross  &  \checkmark  &  \checkmark  &  \checkmark  &  \cross  &  \checkmark  &  \cross \\
  
  &  Ma \etal~\cite{ma2019combination}  &  \checkmark  &  \cross  &  \checkmark  &  \cross  &  \cross  &  \checkmark  &  \cross  &  \checkmark  &  \cross  &  \cross \\
  
 % &  Herath \etal~\cite{herath2022cfgexplainer}  &  \checkmark  &  \cross  &  \checkmark  &  \cross  &  \cross  &  \checkmark  &  \cross  &  \cross  &  \NA  &  \NA \\
  
  &  Aman \etal~\cite{aman2022machine}  &  \cross  &  \checkmark  &  \checkmark  &  \cross  &  \cross  &  \cross  &  \cross  &  \cross  &  \NA  &  \NA\\

\hline
  
\multirow{9}{*}{\textbf{CFA}}  &  C-Flat~\cite{abera2016c}  &  \checkmark  &  \cross  &  \checkmark  &  \cross  &  \checkmark  &  \checkmark  &  \checkmark  &  \checkmark  &  \checkmark  &  \cross \\
  
  &  Atrium~\cite{zeitouni2017atrium}  &  \checkmark  &  \cross  &  \checkmark  &   \checkmark  &  \cross  &  \checkmark  &  \cross  &  \checkmark  &  \checkmark  &  (\checkmark) \\
 
  &  Lo-fat~\cite{dessouky2017fat}  &  \checkmark  &  \cross  &  \checkmark  &   \checkmark  &  \cross  &  \checkmark  &  \cross  &  \checkmark  &  \checkmark  &  (\checkmark) \\
 
  &  Litehax~\cite{dessouky2018litehax}  &  \checkmark  &  \checkmark  &  \checkmark  &  \checkmark  &  \cross  &  \cross  &  \cross  &  \checkmark  &  \checkmark  &  \checkmark\\
 
  &  OAT~\cite{sun2020oat}  &  \checkmark  &  \cross  &  \checkmark  &  \cross  &  \checkmark  &  \checkmark  &  \checkmark  &  \checkmark  &  \checkmark  &  \cross \\
 
  &  Lambda~\cite{kadiyala2020lambda}  &  \cross  &  \cross  &  \checkmark  &  \cross  &  \checkmark  &  \checkmark  &  \cross  &  \cross  &  (\checkmark)  &  \cross \\ % HPC
 
  &  Recfa~\cite{zhang2021recfa}  &  \checkmark  &  \checkmark  &  \checkmark  &  \cross  &  \checkmark  &  \checkmark  &  \checkmark  &  \cross  &  \checkmark  &  \cross \\
 
  &  Tiny-CFA~\cite{nunes2021tiny}  &  \checkmark  &  \checkmark  &  \cross  &  \checkmark  &  \cross  &  \cross  &  \checkmark  &  \cross  &  \checkmark  &  \cross \\
 
  &  BLAST~\cite{yadav2023whole}  &  \checkmark  &  \cross  &  \checkmark  &  \cross  &  \checkmark  &  \cross  &  \checkmark  &  \checkmark  &   \checkmark  &  \cross\\
 
 \bottomrule 
 
  &  \multicolumn{1}{c|}{\textbf{\ourname}}  &  \cross  &  \cross  &  \cross  &  \cross  &  \checkmark  &  \cross  &  \cross  &  \cross  &  \checkmark  &  \checkmark \\

\end{tabular}
}

\label{tab:related_work_assumption}
\end{table*}

CFA can be implemented in various ways: hardware-based solutions, software-based solutions, or a combination thereof.
CFA solutions protect against attacks such as return-oriented programming (ROP) attacks. They either require instrumentation of software as, e.g., C-Flat~\cite{abera2016c}, OAT~\cite{sun2020oat} and Recfa~\cite{zhang2021recfa} or additional hardware components or features as, e.g., Lo-fat~\cite{dessouky2017fat}, Atrium~\cite{zeitouni2017atrium} and Tiny-CFA~\cite{nunes2021tiny}. The approaches that rely on a wide amount of measurements, such as C-Flat~\cite{abera2016c} and BLAST~\cite{yadav2023whole}, send the full execution path or Control-Flow events to the verifier, which has to keep a large database of known benign executions or events. This creates a large computational and storage overhead and the problem of collecting all benign executions or actions. 
Other works rely instead on specific hardware features such as Hardware Performace Counters (HPCs), which provide ample measurements about the execution used as execution's features for the attestation algorithm (e.g., last branch record, branch trace store). Cfimon~\cite{xia2012cfimon} can only detect control-data attacks, but not non-control-data attacks, while Lambda~\cite{kadiyala2020lambda} needs benign programs to be well defined, and only ``large'' deviations from benign control flows are detected. In addition, Numchecker~\cite{wang2013numchecker} focuses on detecting attackers in guest VMs from the host.
More sophisticated approaches such as LiteHAX~\cite{dessouky2018litehax} also track data flows to protect against data-oriented programming (DOP) attacks.

The state-of-the-art is represented by BLAST~\cite{yadav2023whole} which builds on top of C-Flat~\cite{abera2016c} and OAT~\cite{sun2020oat}. BLAST requires access to the complete Control Flow-Graph. The proposed scheme reduces the run-time overhead compared to C-Flat~\cite{abera2016c} by reducing the number of TEE domain switches. However, the approach can only detect control-data attacks (e.g., ROP attacks).

In contrast, our approach, \ourname, does not need a-priory knowledge other than around 10+1 (i.e., the threshold set and \emph{one} trace for training) known-benign execution traces and makes no assumptions about the path coverage of these graphs built from the corresponding traces. \ourname is lightweight, as the model exhibits a minimal number of parameters (cf., \Cref{sec:eval_runtime}, e.g., less than $10,000$) since it does not require keeping a database of known traces.

\subsection{\textbf{ML-based Attestation}}

Machine Learning is utilized for attestation to increase the precision of the attestation mechanism or to eliminate/weaken the strong assumptions of previous works. Therefore, ML is utilized in many stages in an attestation approaches pipeline.

Hu \etal~\cite{hu2019probability} utilizes an ML model to predict the vulnerability of each function in a program. Depending on the prediction, the approach switches between fine and coarse-grained tracing (applying usual attestation hashing methods). While this shortens the execution's duration, their scheme requires access to the software's code and a labeled dataset, including ratings for each function's vulnerabilities. In contrast to our work, time-consuming manual expert labor is needed to create such a dataset.\\
Ma \etal~\cite{ma2019combination} propose a method of attesting Android software. Each program to be attested acquires a Control-Flow Graph (CFG) of the API calls it program executes, which is then fed to a Recurrent Neural Network (RNN) LSTM-based model for prediction. Even though this approach acquires a control-flow graph, the use of RNNs only allows for working on the sequence of API calls, making the whole approach, including the model, resource-heavy. In contrast, \ourname works with GNNs, which are lightweight, allowing fast training and inference time. Further, \ourname only requires \emph{one} execution trace to train the model unsupervised. This way, we can detect all kinds of code-reuse attacks. 

Aman \etal~\cite{aman2022machine} propose to use ML-based classifiers to verify the integrity of an IoT device’s internal state based on its memory contents. The approach converts the memory dump into a grayscale picture where malicious code is detected using regression, decision trees, and Support Vector Machines (SVM). In contrast to our approach, for every attestation, the device's whole memory must be read and converted. By converting binary to a picture representation, possibly information is lost, as it is similar to lossy compression.

\section{Discussion: Real-World Applicability}\label{sec:disc}

We stress that \ourname is feasible in real-world scenarios with reasonable assumptions. In contrast, even the most recent work on CFA, BLAST~\cite{yadav2023whole} only theoretically computes the performance of their approach, assumes knowledge of the critical paths of the CFG and a well-defined, limited set of inputs to the attested software to achieve computation of a complete (relatively to the input set) CFG, which may not be feasible.

A real-world embedded application to be verified through attestation the input set $I$ consist of all possible input combinations. The cardinality of such a set is exponential and can be written as $|I| \to \infty$.
When we deploy an attestation scheme that requires access to a CCFG (cf., Definition \ref{eq:ccfg}), we are practically limiting the combinations of the input set, which can be written as $S \subset I$ where $S$ is the limited set of inputs, thus, limiting the attestation capability or the facility of the system.
Specifically, the system could be in a benign state recognized as malicious as that specific input was not initially in the set of inputs. %This will lead to a false positive. 
Additionally, restricting the system to specific inputs may hinder the system to work correctly when faced with edge cases, specifically all $i \notin S$, where $i$ is one input.
Moreover, manual labeling and, therefore, only attesting the critical paths of software may not be sufficient, as DOP attacks may not be detected or ROP attacks can, e.g., open a shell in a non-critical path. Therefore, the attestation of the full software, including its used libraries, is necessary.

By contrast, \ourname does not require such prior measurements or critical path labeling, circumventing the above-mentioned constraints. As we described in Definition \ref{def:embs}, we leverage a geometric deep learning model for extracting latent features of execution traces that maintain a strict correspondence between these data. This allows \ourname to achieve attestation of the full software through the learning of a benign behavior, overcoming the limitations of related work.

\section{Conclusion}\label{sec:concl}
 In this paper, we present a new method for Control-Flow Attestation (CFA) in embedded systems and IoT devices. Our approach addresses the limitations of existing CFA techniques that make them challenging to be implemented in real-world scenarios. These limitations include the requirement of a large set of known-benign executions, access to a complete Control-Flow Graph (CFG), and information about the system's internal state, which may not be feasible, particularly for IP-protected code. In contrast, \ourname detects Code Reuse Attacks (CRAs) without requiring any information except for the CPU's program counter, is lightweight, and therefore applicable to resource-constrained devices. \ourname uses Unsupervised Graph Neural Networks (GNNs) to analyze the control flow trace of only \emph{one} execution, allowing it to detect both return-oriented programming (ROP) and data-oriented programming (DOP) attacks.
 The results of our experiments demonstrate the efficacy of \ourname in detecting CRAs in real-world attacks in real-world scenarios, proving it suitable for resource-constrained embedded devices running IP-protected code. \ourname can detect ROP and DOP attacks with an F1-Score of on average 97.49\% and 84.42\%, respectively, while also maintaining a low False-Positive Rate of on average 5.47\%, for OpenSSL. Similarly, \ourname achieved on average 98.03\% (ROP) and 91.01\% (DOP) F1-Score and 3.19\% FPR for Embench. This makes our approach a promising solution for protecting the integrity and authenticity of embedded systems and IoT devices.

\section*{Acknowledgements.} 
\footnotesize{
Our research work was partially funded by Deutsche Forschungsgemeinschaft (DFG) – SFB 1119 – 236615297, the European Union under Horizon Europe Programme – Grant Agreement 101070537 – CrossCon.}

\clearpage

%-------------------------------------------------------------------------------

\bibliographystyle{IEEEtran}
\bibliography{IEEEabrv, bibliography.bib}

\begin{thebibliography}{10}
\providecommand{\url}[1]{#1}
\csname url@samestyle\endcsname
\providecommand{\newblock}{\relax}
\providecommand{\bibinfo}[2]{#2}
\providecommand{\BIBentrySTDinterwordspacing}{\spaceskip=0pt\relax}
\providecommand{\BIBentryALTinterwordstretchfactor}{4}
\providecommand{\BIBentryALTinterwordspacing}{\spaceskip=\fontdimen2\font plus
\BIBentryALTinterwordstretchfactor\fontdimen3\font minus \fontdimen4\font\relax}
\providecommand{\BIBforeignlanguage}[2]{{%
\expandafter\ifx\csname l@#1\endcsname\relax
\typeout{** WARNING: IEEEtran.bst: No hyphenation pattern has been}%
\typeout{** loaded for the language `#1'. Using the pattern for}%
\typeout{** the default language instead.}%
\else
\language=\csname l@#1\endcsname
\fi
#2}}
\providecommand{\BIBdecl}{\relax}
\BIBdecl

\bibitem{seshadri2004swatt}
A.~Seshadri, A.~Perrig, L.~Van~Doorn, and P.~Khosla, ``Swatt: Software-based attestation for embedded devices,'' in \emph{IEEE Symposium on Security and Privacy, 2004. Proceedings. 2004}.\hskip 1em plus 0.5em minus 0.4em\relax IEEE, 2004, pp. 272--282.

\bibitem{seshadri2006scuba}
A.~Seshadri, M.~Luk, A.~Perrig, L.~Van~Doorn, and P.~Khosla, ``Scuba: Secure code update by attestation in sensor networks,'' in \emph{Proceedings of the 5th ACM workshop on Wireless security}, 2006, pp. 85--94.

\bibitem{chen2017secure}
B.~Chen, X.~Dong, G.~Bai, S.~Jauhar, and Y.~Cheng, ``Secure and efficient software-based attestation for industrial control devices with arm processors,'' in \emph{Proceedings of the 33rd Annual Computer Security Applications Conference}, 2017, pp. 425--436.

\bibitem{seshadri2005pioneer}
A.~Seshadri, M.~Luk, E.~Shi, A.~Perrig, L.~Van~Doorn, and P.~Khosla, ``Pioneer: verifying code integrity and enforcing untampered code execution on legacy systems,'' in \emph{Proceedings of the twentieth ACM symposium on Operating systems principles}, 2005, pp. 1--16.

\bibitem{li2011viper}
Y.~Li, J.~M. McCune, and A.~Perrig, ``Viper: Verifying the integrity of peripherals' firmware,'' in \emph{Proceedings of the 18th ACM conference on Computer and communications security}, 2011, pp. 3--16.

\bibitem{seshadri2008sake}
A.~Seshadri, M.~Luk, and A.~Perrig, ``Sake: Software attestation for key establishment in sensor networks,'' in \emph{International Conference on Distributed Computing in Sensor Systems}.\hskip 1em plus 0.5em minus 0.4em\relax Springer, 2008, pp. 372--385.

\bibitem{krauss2007detecting}
C.~Krau{\ss}, F.~Stumpf, and C.~Eckert, ``Detecting node compromise in hybrid wireless sensor networks using attestation techniques,'' in \emph{European Workshop on Security in Ad-hoc and Sensor Networks}.\hskip 1em plus 0.5em minus 0.4em\relax Springer, 2007, pp. 203--217.

\bibitem{agrawal2015program}
S.~Agrawal, M.~L. Das, A.~Mathuria, and S.~Srivastava, ``Program integrity verification for detecting node capture attack in wireless sensor network,'' in \emph{International Conference on Information Systems Security}.\hskip 1em plus 0.5em minus 0.4em\relax Springer, 2015, pp. 419--440.

\bibitem{belkin}
T.~Post, ``Belkin iot smart plug flaw allows remote code execution in smart homes,'' \url{https://threatpost.com/belkin-iot-smart-plug-flaw-allows-remote-code-execution-in-smart-homes/136732/}, 2018.

\bibitem{modems}
Arstechnica, ``Exploit that gives remote access affects ~200 million cable modems,'' \url{https://arstechnica.com/information-technology/2020/01/exploit-that-gives-remote-access-affects-200-million-cable-modems/}, 2020.

\bibitem{firewall}
{Zd Net}, ``Nasty linux netfilter firewall security hole found,'' \url{https://www.zdnet.com/article/nasty-linux-netfilter-firewall-security-hole-found/}, 2022.

\bibitem{samsung}
Bleepingcomputer, ``Cisa warns of samsung aslr bypass flaw exploited in attacks,'' \url{https://www.bleepingcomputer.com/news/security/cisa-warns-of-samsung-aslr-bypass-flaw-exploited-in-attacks/}, 2023.

\bibitem{abera2016c}
N.~Asokan, L.~Davi, J.-E. Ekberg, T.~Nyman, A.~Paverd, A.-R. Sadeghi, and G.~Tsudik, ``C-flat: control-flow attestation for embedded systems software,'' in \emph{Proceedings of the 2016 ACM SIGSAC Conference on Computer and Communications Security}, 2016, pp. 743--754.

\bibitem{zeitouni2017atrium}
S.~Zeitouni, G.~Dessouky, O.~Arias, D.~Sullivan, A.~Ibrahim, Y.~Jin, and A.-R. Sadeghi, ``Atrium: Runtime attestation resilient under memory attacks,'' in \emph{2017 IEEE/ACM International Conference on Computer-Aided Design (ICCAD)}.\hskip 1em plus 0.5em minus 0.4em\relax IEEE, 2017, pp. 384--391.

\bibitem{dessouky2017fat}
G.~Dessouky, S.~Zeitouni, T.~Nyman, A.~Paverd, L.~Davi, P.~Koeberl, N.~Asokan, and A.-R. Sadeghi, ``Lo-fat: Low-overhead control flow attestation in hardware,'' in \emph{Proceedings of the 54th Annual Design Automation Conference 2017}, 2017, pp. 1--6.

\bibitem{dessouky2018litehax}
G.~Dessouky, T.~Abera, A.~Ibrahim, and A.-R. Sadeghi, ``Litehax: lightweight hardware-assisted attestation of program execution,'' in \emph{2018 IEEE/ACM International Conference on Computer-Aided Design (ICCAD)}.\hskip 1em plus 0.5em minus 0.4em\relax IEEE, 2018, pp. 1--8.

\bibitem{sun2020oat}
Z.~Sun, B.~Feng, L.~Lu, and S.~Jha, ``Oat: Attesting operation integrity of embedded devices,'' in \emph{2020 IEEE Symposium on Security and Privacy (SP)}.\hskip 1em plus 0.5em minus 0.4em\relax IEEE, 2020, pp. 1433--1449.

\bibitem{zhang2021recfa}
Y.~Zhang, X.~Liu, C.~Sun, D.~Zeng, G.~Tan, X.~Kan, and S.~Ma, ``Recfa: Resilient control-flow attestation,'' in \emph{Annual Computer Security Applications Conference}, 2021, pp. 311--322.

\bibitem{yadav2023whole}
N.~Yadav and V.~Ganapthy, ``Whole-program control-flow path attestation,'' in \emph{30th ACM conference on Computer and Communications Security}, 2023, early available at: \url{https://www.csa.iisc.ac.in/~vg/papers/ccs2023/}.

\bibitem{nunes2021tiny}
I.~D.~O. Nunes, S.~Jakkamsetti, and G.~Tsudik, ``Tiny-cfa: Minimalistic control-flow attestation using verified proofs of execution,'' in \emph{2021 Design, Automation \& Test in Europe Conference \& Exhibition (DATE)}.\hskip 1em plus 0.5em minus 0.4em\relax IEEE, 2021, pp. 641--646.

\bibitem{conrado2023bounded}
G.~K. Conrado, A.~Goharshady, and C.~K. Lam, ``The bounded pathwidth of control-flow graphs,'' in \emph{ACM Conference on Object-Oriented Programming, Systems, Languages, and Applications, OOPSLA 2023}, 2023.

\bibitem{frassetto2022cfinsight}
T.~Frassetto, P.~Jauernig, D.~Koisser, and A.-R. Sadeghi, ``Cfinsight: A comprehensive metric for cfi policies,'' in \emph{29th Annual Network and Distributed System Security Symposium}.\hskip 1em plus 0.5em minus 0.4em\relax NDSS, 2022.

\bibitem{theiling2002control}
H.~Theiling, ``Control flow graphs for real-time systems analysis: reconstruction from binary executables and usage in ilp-based path analysis,'' PhD thesis, Saarland University, 2002.

\bibitem{van2007relating}
D.~Van~Horn and H.~G. Mairson, ``Relating complexity and precision in control flow analysis,'' \emph{ACM SIGPLAN Notices}, vol.~42, no.~9, pp. 85--96, 2007.

\bibitem{xu2009constructing}
L.~Xu, F.~Sun, and Z.~Su, ``Constructing precise control flow graphs from binaries,'' \emph{University of California, Davis, Tech. Rep}, pp. 14--23, 2009.

\bibitem{zhu2021constructing}
K.~Zhu, Y.~Lu, H.~Huang, L.~Yu, and J.~Zhao, ``Constructing more complete control flow graphs utilizing directed gray-box fuzzing,'' \emph{Applied Sciences}, vol.~11, no.~3, p. 1351, 2021.

\bibitem{hu2019probability}
J.~Hu, D.~Huo, M.~Wang, Y.~Wang, Y.~Zhang, and Y.~Li, ``A probability prediction based mutable control-flow attestation scheme on embedded platforms,'' in \emph{2019 18th IEEE International Conference On Trust, Security And Privacy In Computing And Communications/13th IEEE International Conference On Big Data Science And Engineering (TrustCom/BigDataSE)}.\hskip 1em plus 0.5em minus 0.4em\relax IEEE, 2019, pp. 530--537.

\bibitem{ma2019combination}
Z.~Ma, H.~Ge, Y.~Liu, M.~Zhao, and J.~Ma, ``A combination method for android malware detection based on control flow graphs and machine learning algorithms,'' \emph{IEEE access}, vol.~7, pp. 21\,235--21\,245, 2019.

\bibitem{aman2022machine}
M.~N. Aman, H.~Basheer, J.~W. Wong, J.~Xu, H.~W. Lim, and B.~Sikdar, ``Machine learning based attestation for the internet of things using memory traces,'' \emph{IEEE Internet of Things Journal}, 2022.

\bibitem{wilander.ripe}
J.~Wilander, N.~Nikiforakis, Y.~Younan, M.~Kamkar, and W.~Joosen, ``{RIPE}: Runtime intrusion prevention evaluator,'' in \emph{In Proceedings of the 27th Annual Computer Security Applications Conference, {ACSAC}}.\hskip 1em plus 0.5em minus 0.4em\relax ACM, 2011.

\bibitem{bennett2022embench}
J.~Bennett, P.~Dabbelt, C.~Garlati, G.~Madhusudan, T.~Mudge, and D.~Patterson, ``Embench: An evolving benchmark suite for embedded iot computers from an academic-industrial cooperative,'' 2022.

\bibitem{turing1936computable}
A.~M. Turing \emph{et~al.}, ``On computable numbers, with an application to the entscheidungsproblem,'' \emph{J. of Math}, vol.~58, no. 345-363, p.~5, 1936.

\bibitem{granata2013maximum}
D.~Granata, R.~Cerulli, M.~G. Scutella, A.~Raiconi \emph{et~al.}, ``Maximum flow problems and an np-complete variant on edge-labeled graphs,'' \emph{Handbook of combinatorial optimization}, pp. 1913--1948, 2013.

\bibitem{rimsa2021practical}
A.~Rimsa, J.~Nelson~Amaral, and F.~M. Pereira, ``Practical dynamic reconstruction of control flow graphs,'' \emph{Software: Practice and Experience}, vol.~51, no.~2, pp. 353--384, 2021.

\bibitem{shacham2007geometry}
H.~Shacham, ``The geometry of innocent flesh on the bone: Return-into-libc without function calls (on the x86),'' in \emph{Proceedings of the 14th ACM conference on Computer and communications security}, 2007, pp. 552--561.

\bibitem{bletsch2011jump}
T.~Bletsch, X.~Jiang, V.~W. Freeh, and Z.~Liang, ``Jump-oriented programming: a new class of code-reuse attack,'' in \emph{Proceedings of the 6th ACM Symposium on Information, Computer and Communications Security}, 2011, pp. 30--40.

\bibitem{checkoway2010return}
S.~Checkoway, L.~Davi, A.~Dmitrienko, A.-R. Sadeghi, H.~Shacham, and M.~Winandy, ``Return-oriented programming without returns,'' in \emph{Proceedings of the 17th ACM conference on Computer and communications security}, 2010, pp. 559--572.

\bibitem{ispoglou2018block}
K.~K. Ispoglou, B.~AlBassam, T.~Jaeger, and M.~Payer, ``Block oriented programming: Automating data-only attacks,'' in \emph{Proceedings of the 2018 ACM SIGSAC Conference on Computer and Communications Security}, 2018, pp. 1868--1882.

\bibitem{hu2016data}
H.~Hu, S.~Shinde, S.~Adrian, Z.~L. Chua, P.~Saxena, and Z.~Liang, ``Data-oriented programming: On the expressiveness of non-control data attacks,'' in \emph{2016 IEEE Symposium on Security and Privacy (SP)}.\hskip 1em plus 0.5em minus 0.4em\relax IEEE, 2016, pp. 969--986.

\bibitem{sperduti1997supervised}
A.~Sperduti and A.~Starita, ``Supervised neural networks for the classification of structures,'' \emph{IEEE Transactions on Neural Networks}, vol.~8, no.~3, pp. 714--735, 1997.

\bibitem{welling2016semi}
M.~Welling and T.~N. Kipf, ``Semi-supervised classification with graph convolutional networks,'' in \emph{J. International Conference on Learning Representations (ICLR 2017)}, 2016.

\bibitem{kipf2016variational}
T.~N. Kipf and M.~Welling, ``Variational graph auto-encoders,'' \emph{arXiv preprint arXiv:1611.07308}, 2016.

\bibitem{kingma2013auto}
D.~P. Kingma and M.~Welling, ``Auto-encoding variational bayes,'' \emph{arXiv preprint arXiv:1312.6114}, 2013.

\bibitem{xu2021understanding}
M.~Xu, ``Understanding graph embedding methods and their applications,'' \emph{SIAM Review}, vol.~63, no.~4, pp. 825--853, 2021.

\bibitem{molnar2006program}
D.~Molnar, M.~Piotrowski, D.~Schultz, and D.~Wagner, ``The program counter security model: Automatic detection and removal of control-flow side channel attacks,'' in \emph{Information Security and Cryptology-ICISC 2005: 8th International Conference, Seoul, Korea, December 1-2, 2005, Revised Selected Papers 8}.\hskip 1em plus 0.5em minus 0.4em\relax Springer, 2006, pp. 156--168.

\bibitem{hausdorff1914grundzuge}
F.~Hausdorff, \emph{Grundz{\"u}ge der mengenlehre}.\hskip 1em plus 0.5em minus 0.4em\relax von Veit, 1914, vol.~7.

\bibitem{dynamorio:phd_thesis}
D.~Bruening and S.~Amarasinghe, ``Efficient, transparent, and comprehensive runtime code manipulation,'' Ph.D. dissertation, Massachusetts Institute of Technology, Department of Electrical Engineering, 2004.

\bibitem{eldefrawy2012smart}
K.~Eldefrawy, G.~Tsudik, A.~Francillon, and D.~Perito, ``Smart: secure and minimal architecture for (establishing dynamic) root of trust.'' in \emph{Ndss}, vol.~12, 2012, pp. 1--15.

\bibitem{koeberl2014trustlite}
P.~Koeberl, S.~Schulz, A.-R. Sadeghi, and V.~Varadharajan, ``Trustlite: A security architecture for tiny embedded devices,'' in \emph{Proceedings of the Ninth European Conference on Computer Systems}, 2014, pp. 1--14.

\bibitem{brasser2015tytan}
F.~Brasser, B.~El~Mahjoub, A.-R. Sadeghi, C.~Wachsmann, and P.~Koeberl, ``Tytan: Tiny trust anchor for tiny devices,'' in \emph{Proceedings of the 52nd annual design automation conference}, 2015, pp. 1--6.

\bibitem{carpent2018remote}
X.~Carpent, N.~Rattanavipanon, and G.~Tsudik, ``Remote attestation of iot devices via smarm: Shuffled measurements against roving malware,'' in \emph{2018 IEEE international symposium on hardware oriented security and trust (HOST)}.\hskip 1em plus 0.5em minus 0.4em\relax IEEE, 2018, pp. 9--16.

\bibitem{aman2018att}
M.~N. Aman and B.~Sikdar, ``Att-auth: A hybrid protocol for industrial iot attestation with authentication,'' \emph{IEEE Internet of Things Journal}, vol.~5, no.~6, pp. 5119--5131, 2018.

\bibitem{aman2021prom}
M.~N. Aman, M.~H. Basheer, S.~Dash, A.~Sancheti, J.~W. Wong, J.~Xu, H.~W. Lim, and B.~Sikdar, ``Prom: Passive remote attestation against roving malware in multicore iot devices,'' \emph{IEEE Systems Journal}, vol.~16, no.~1, pp. 789--800, 2021.

\bibitem{wang2013numchecker}
X.~Wang and R.~Karri, ``Numchecker: Detecting kernel control-flow modifying rootkits by using hardware performance counters,'' in \emph{Proceedings of the 50th Annual Design Automation Conference}, 2013, pp. 1--7.

\bibitem{xia2012cfimon}
Y.~Xia, Y.~Liu, H.~Chen, and B.~Zang, ``Cfimon: Detecting violation of control flow integrity using performance counters,'' in \emph{IEEE/IFIP International Conference on Dependable Systems and Networks (DSN 2012)}.\hskip 1em plus 0.5em minus 0.4em\relax IEEE, 2012, pp. 1--12.

\bibitem{kadiyala2020lambda}
S.~P. Kadiyala, M.~Alam, Y.~Shrivastava, S.~Patranabis, M.~F.~B. Abbas, A.~K. Biswas, D.~Mukhopadhyay, and T.~Srikanthan, ``Lambda: Lightweight assessment of malware for embedded architectures,'' \emph{ACM Transactions on Embedded Computing Systems (TECS)}, vol.~19, no.~4, pp. 1--31, 2020.

\end{thebibliography}

\clearpage
\appendix
\begin{appendices}

\section{Features Extracted}
In \Cref{tab:features_dop_rop} we report the features we extract while preprocessing an execution into a graph.
\begin{table}[h!]
\centering
\caption{For each extracted feature, we have specified the specific detection purpose for which it is intended.}

\begin{tabular}{c|lc}
\textbf{\#}&\multicolumn{1}{l}{\textbf{Feature}} & \textbf{Use} \\
\hline
1&Vertex degree                         &   ROP  \\
2&Number of visits                      &   ROP \& DOP           \\
3&First visit                           &   ROP \& DOP           \\
4&Last visit                            &   ROP \& DOP           \\
5&Incoming edges                        &   ROP           \\
6&Outgoing edges                        &   ROP          \\
7&Frequency of visits  & ROP \& DOP \\
8&Time of use & ROP \& DOP \\
9&Std visits & ROP \& DOP\\
10&Mean dist. visits & ROP \& DOP\\
11&Mean visits & ROP \& DOP\\
12&Mean N. visits. In. Neigh. & ROP \& DOP\\
13&Mean N. visits. Out. Neigh. & ROP \& DOP\\
14&Mean last visit In. Neigh. & ROP \& DOP\\
15&Mean last visit Out. Neigh. & ROP \& DOP\\
\end{tabular}

\label{tab:features_dop_rop}
\end{table}

\section{Evaluation Results}
\begin{figure*}[ht]
    \centering
    \subfigure[Diffie-Hellman]{\includegraphics[width=0.19\linewidth,clip]{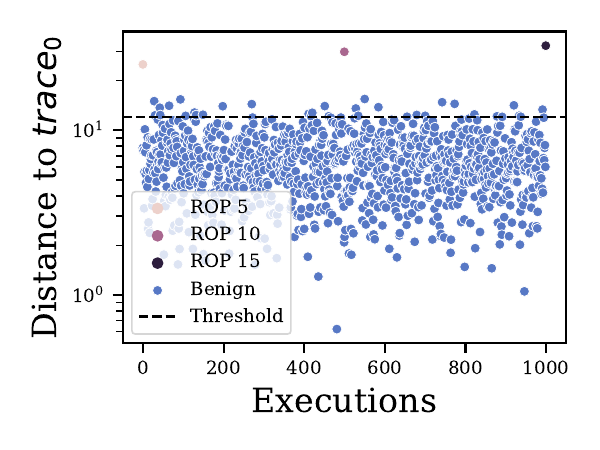}} 
    \subfigure[DES]{\includegraphics[width=0.19\linewidth,clip]{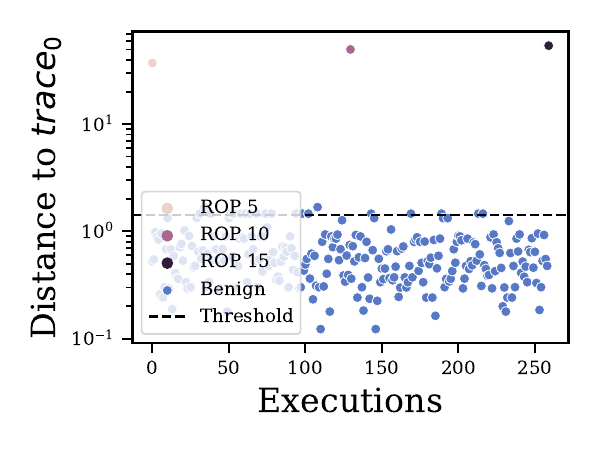}} 
    \subfigure[DESX]{\includegraphics[width=0.19\linewidth,clip]{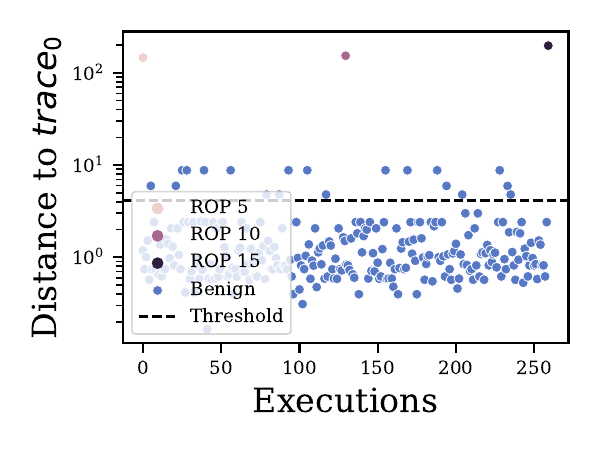}}
    \subfigure[GOST]{\includegraphics[width=0.19\linewidth,clip]{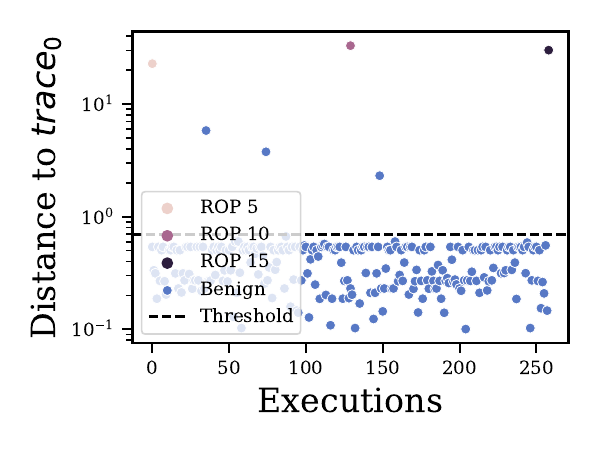}}
    \subfigure[AES]{\includegraphics[width=0.19\linewidth,clip]{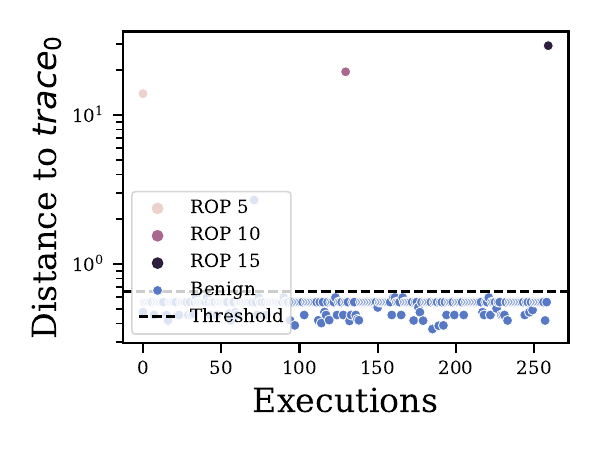}}
    \caption{Evaluation of ROP attacks. In the label ``ROP $x$", $x$ indicates the length of the ROP attack in the trace. Each data point represents the mean of 50 evaluated ROP traces of the same kind. For better readability, the presented data points are capped at 15, as higher data points are naturally detected as malicious.}
    \label{fig:rop_eval}
\end{figure*}

\begin{figure*}[ht]
    \centering
    \subfigure[Diffie-Hellman]{\includegraphics[width=0.19\linewidth,clip]{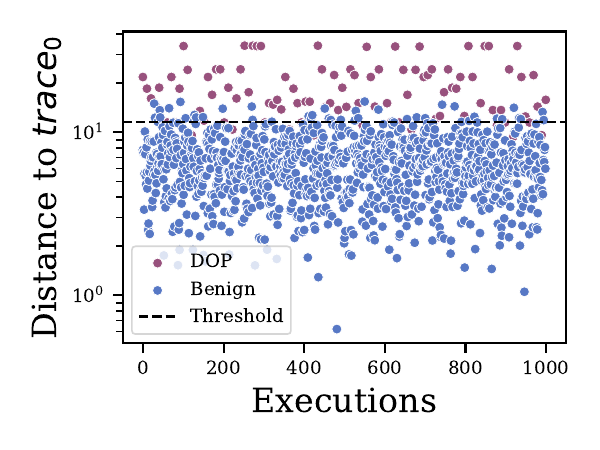}} 
    \subfigure[DES]{\includegraphics[width=0.19\linewidth,clip]{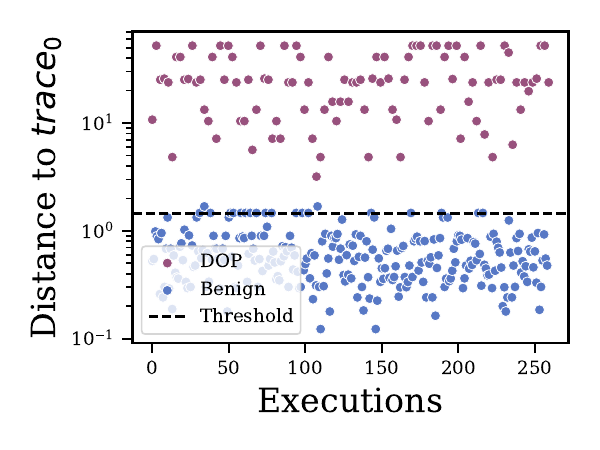}} 
    \subfigure[DESX]{\includegraphics[width=0.19\linewidth,clip]{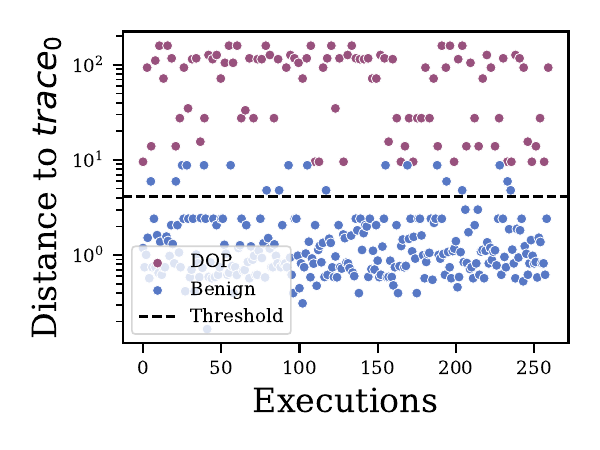}}
    \subfigure[GOST]{\includegraphics[width=0.19\linewidth,clip]{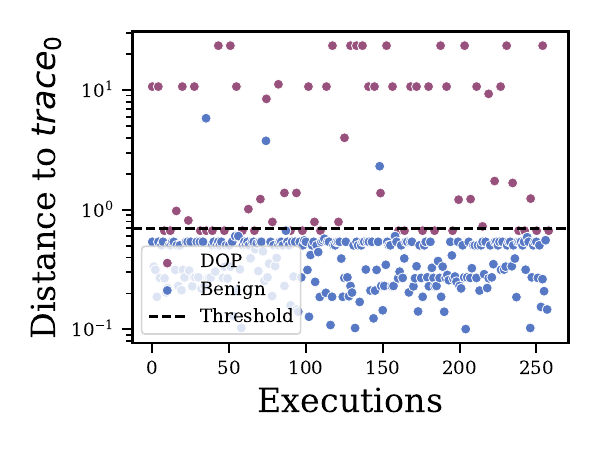}}
    \subfigure[AES]{\includegraphics[width=0.19\linewidth,clip]{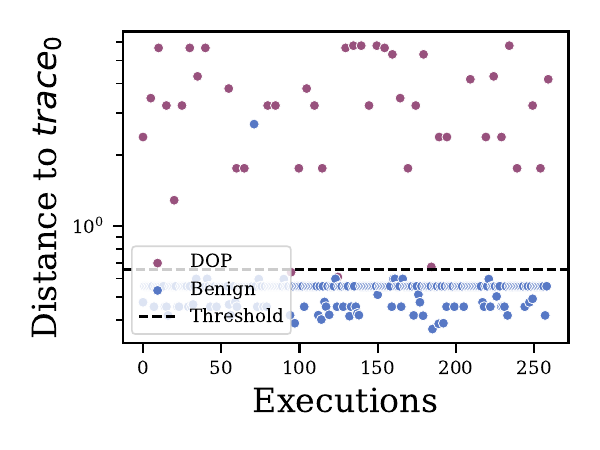}}
    \caption{Evaluation of DOP attacks. For each software, 100 different DOP attack traces have been generated, except for AES, where we generated 53 attacks.}
    \label{fig:dop_eval}
\end{figure*}

In \Cref{fig:rop_eval} and \Cref{fig:dop_eval} we report the evaluation results in a graphic manner for the cryptographic dataset.

\section{Real-World Attacks Results}\label{app:ripe}
\begin{figure*}[ht]
	\centering
	\includegraphics[scale=0.4,trim={0em 0em 0em 0},clip]{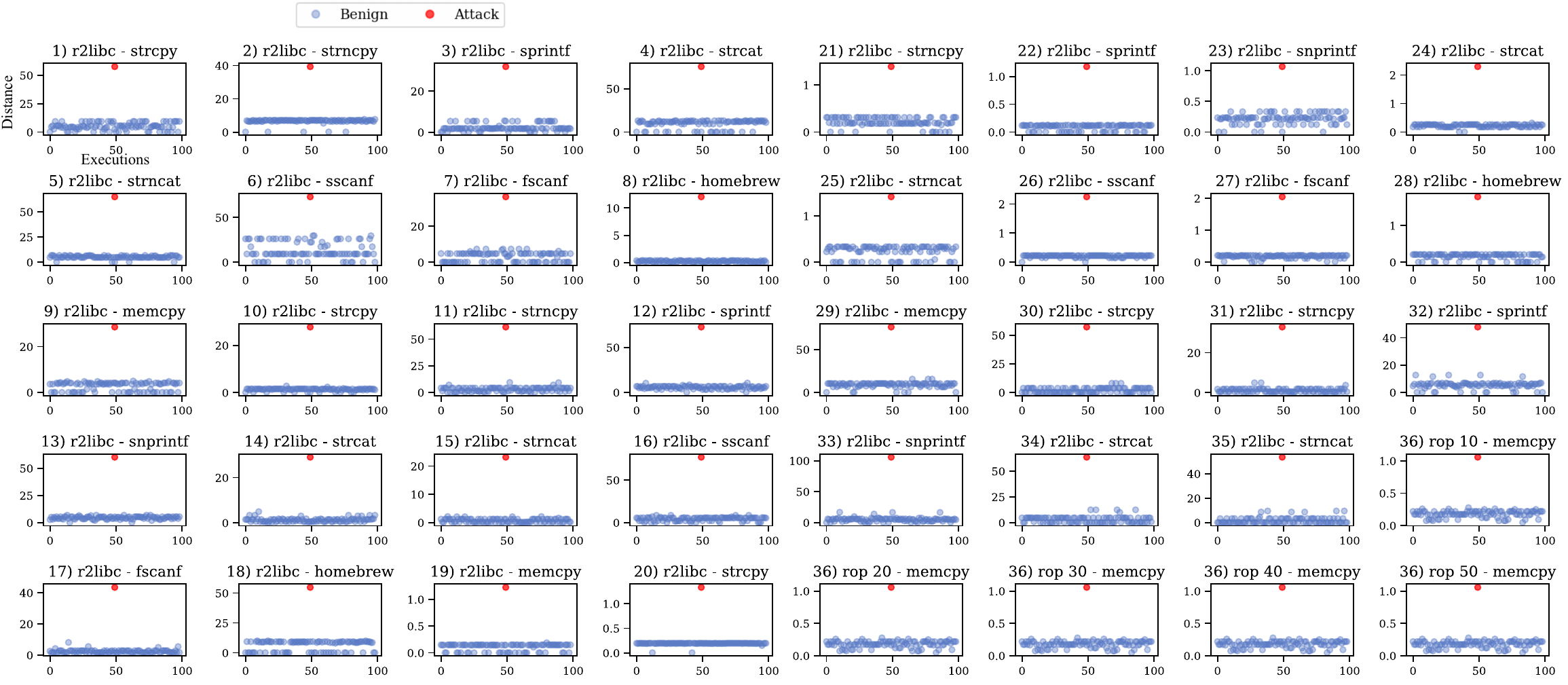}
	\caption{Evaluation of real-world attacks performed on the picojpg binary of Embench. The blue and red dots represent, respectively, the benign executions and the performed attack. In each title is specified the kind of attack and the methodology. When the attack is a ROP attack, it is reported the attack chain length as "ROP $x$", where $x$ is the length. The x-axis enumerates the executions, the y-axis represents the distance to the training trace. All the details on the attacks are reported in \Cref{tab:ripe_details}} 
	\label{fig:RIPE}
\end{figure*}

In \Cref{fig:RIPE} we report the evaluation of real-world attacks obtained through the use of the framework RIPE. The details of each attack are reported in \Cref{tab:ripe_details}.

\begin{table}[ht]
\centering
\caption{Real-world attacks details. ROP* indicates that the ROP attack setting described in the table are valid for the 5 ROP attack performed with different chain lengths.}
\resizebox{\columnwidth}{!}{

\begin{tabular}{c|ccccc}
Bin. \# & Attack Type & Overflow Type & Attack Location & Target Code Ptr. & Function Exploited\\
\hline
1 & r2libc & direct & stack & funcptrstackvar & strcpy\\
2 & r2libc & direct & stack & funcptrstackvar & strncpy\\
3 & r2libc & direct & stack & funcptrstackvar & sprintf\\
4 & r2libc & direct & stack & funcptrstackvar & strcat\\
5 & r2libc & direct & stack & funcptrstackvar & strncat\\
6 & r2libc & direct & stack & funcptrstackvar & sscanf\\
7 & r2libc & direct & stack & funcptrstackvar & fscanf\\
8 & r2libc & direct & stack & funcptrstackvar & homebrew\\
9 & r2libc & direct & stack & structfuncptrstack & memcpy\\
10 & r2libc & direct & stack & structfuncptrstack & strcpy\\
11 & r2libc & direct & stack & structfuncptrstack & strncpy\\
12 & r2libc & direct & stack & structfuncptrstack & sprintf\\
13 & r2libc & direct & stack & structfuncptrstack & snprintf\\
14 & r2libc & direct & stack & structfuncptrstack & strcat\\
15 & r2libc & direct & stack & structfuncptrstack & strncat\\
16 & r2libc & direct & stack & structfuncptrstack & sscanf\\
17 & r2libc & direct & stack & structfuncptrstack & fscanf\\
18 & r2libc & direct & stack & structfuncptrstack & homebrew\\
19 & r2libc & direct & heap & funcptrheap & memcpy\\
20 & r2libc & direct & heap & funcptrheap & strcpy\\
21 & r2libc & direct & heap & funcptrheap & strncpy\\
22 & r2libc & direct & heap & funcptrheap & sprintf\\
23 & r2libc & direct & heap & funcptrheap & snprintf\\
24 & r2libc & direct & heap & funcptrheap & strcat\\
25 & r2libc & direct & heap & funcptrheap & strncat\\
26 & r2libc & direct & heap & funcptrheap & sscanf\\
27 & r2libc & direct & heap & funcptrheap & fscanf\\
28 & r2libc & direct & heap & funcptrheap & homebrew\\
29 & r2libc & direct & heap & structfuncptrheap & memcpy\\
30 & r2libc & direct & heap & structfuncptrheap & strcpy\\
31 & r2libc & direct & heap & structfuncptrheap & strncpy\\
32 & r2libc & direct & heap & structfuncptrheap & sprintf\\
33 & r2libc & direct & heap & structfuncptrheap & snprintf\\
34 & r2libc & direct & heap & structfuncptrheap & strcat\\
35 & r2libc & direct & heap & structfuncptrheap & strncat\\
36 & ROP* & indirect & stack & baseptr & memcpy\\
\end{tabular}

}
\label{tab:ripe_details}
\end{table}

\section{Threshold Ablation Study}\label{app:threshold_ablation_study}

\begin{figure*}[ht]
    \centering
    \subfigure[Cryptographic Dataset]{\includegraphics[width=0.48\linewidth,clip]{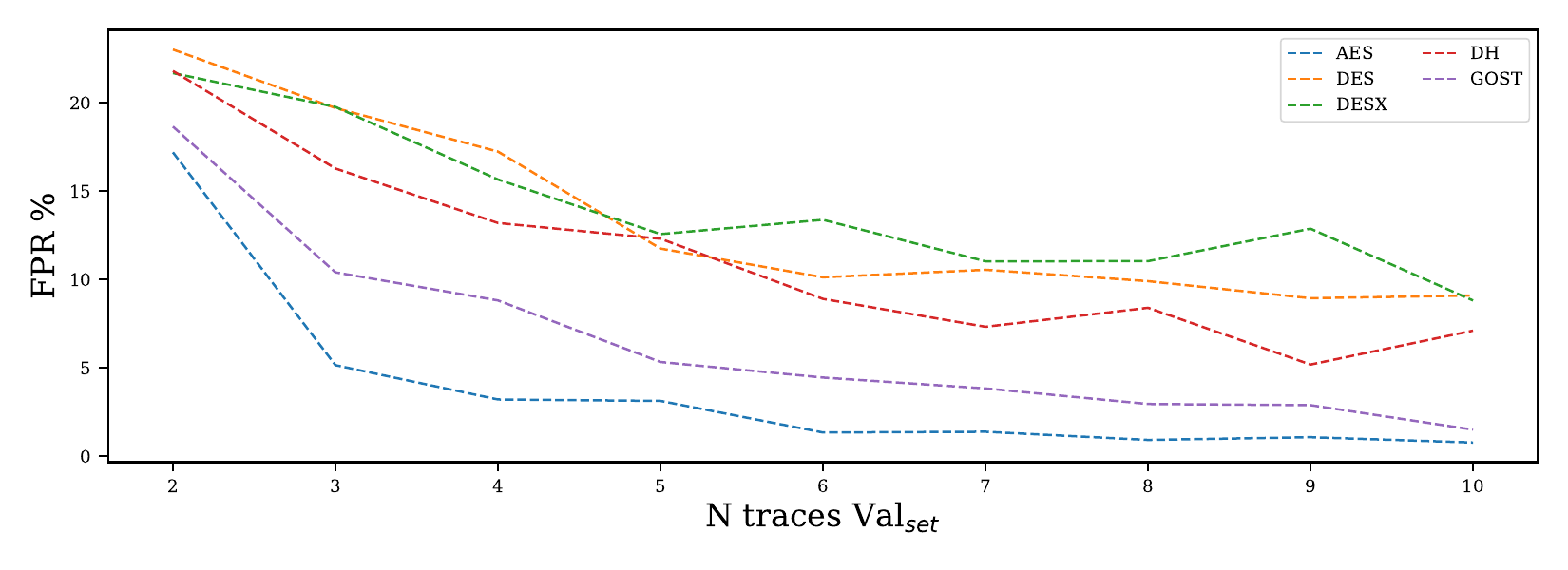}} 
    \subfigure[Embench Dataset]{\includegraphics[width=0.48\linewidth,clip]{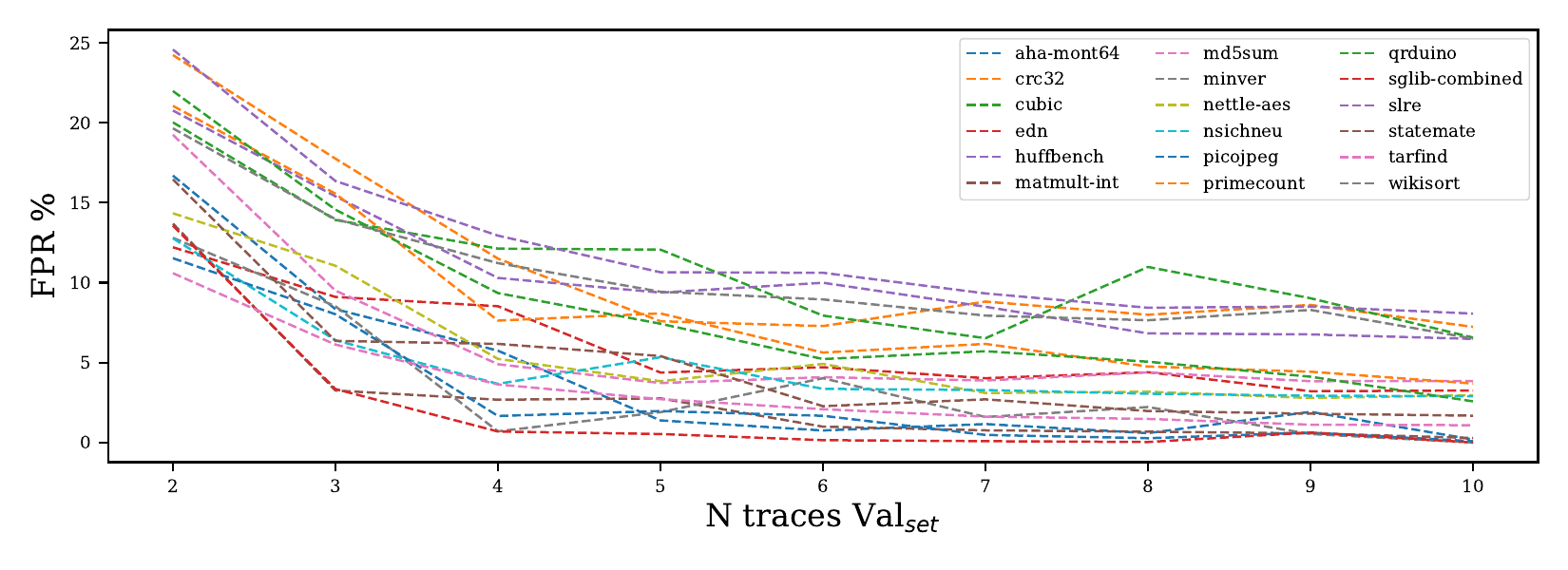}} 
    \caption{Ablation study of the number of traces used for computing the attestation threshold, we observe the change of False-Positive-Rate. The experiments are repeated 100 times.}
    \label{fig:ablation_fpr}
\end{figure*}

\Cref{fig:ablation_fpr} reports the performance of \ourname in terms of the false-positive-rate when considering a different amount of traces for computing the attestation threshold. In \Cref{tab:ROP_ablation} and \Cref{tab:DOP_ablation} we report the detailed results for three different amounts of used traces.

\begin{table*}[ht]
\centering
\caption{Ablation study of the number of traces used for computing the attestation threshold. The results are about the detection of ROP and DOP attacks. The experiments are repeated 100 times. For brevity, we report only the selection 2, 5 and 10 traces. All the values are in percentages.
}
\resizebox{\linewidth}{!}{
\begin{tabular}{lcccc|cccc|cccc|ccc|ccc|ccc}
& \multicolumn{12}{c|}{\textbf{ROP}} & \multicolumn{9}{c}{\textbf{DOP}} \\

 & \multicolumn{4}{c|}{\textbf{n=2}} & \multicolumn{4}{c|}{\textbf{n=5}} & \multicolumn{4}{c|}{\textbf{n=10}} & \multicolumn{3}{c|}{\textbf{n=2}} & \multicolumn{3}{c|}{\textbf{n=5}} & \multicolumn{3}{c}{\textbf{n=10}} \\
 
 & \textbf{FPR} & \textbf{Re.} & \textbf{Pr.} & \textbf{F1-Score} & \textbf{FPR} & \textbf{Re.} & \textbf{Pr.} & \textbf{F1-Score} & \textbf{FPR} & \textbf{Re.} & \textbf{Pr.} & \textbf{F1-Score} & \textbf{Re.} & \textbf{Pr.} & \textbf{F1-Score} & \textbf{Re.} & \textbf{Pr.} & \textbf{F1-Score} & \textbf{Re.} & \textbf{Pr.} & \textbf{F1-Score} \\ \hline
\multicolumn{1}{l|}{\textbf{DH}} & 21.81 & 97.18 & 79.06 & 85.8 & 12.32 & 96.28 & 85.78 & 89.98 & 7.12 & 95.44 & 90.59 & 92.58  & 90.37 & 46.77 & 54.47 & 87.48 & 55.68 & 61.87 & 86.03 & 62.51 & 69.22 \\
\multicolumn{1}{l|}{\textbf{AES}} & 17.2 & 99.55 & 94.17 & 96.63 & 3.139 & 99.22 & 98.84 & 99.01 & 0.78 & 99.06 & 99.69 & 99.37  & 85.79 & 67.35 & 70.85 & 80.25 & 89.43 & 83.47 & 78.92 & 95.88 & 86.41 \\
\multicolumn{1}{l|}{\textbf{DES}} & 23.02 & 99.86 & 92.15 & 95.7 & 11.76 & 99.75 & 95.72 & 97.64 & 9.10 & 99.7 & 96.58 & 98.09  & 100.0 & 69.78 & 80.31 & 100.0 & 80.43 & 88.21 & 100.0 & 82.79 & 90.1 \\
\multicolumn{1}{l|}{\textbf{DESX}} & 21.69 & 99.45 & 92.47 & 95.71 & 12.58 & 99.37 & 95.4 & 97.29 & 8.82 & 99.3 & 96.66 & 97.95  & 97.92 & 70.01 & 79.78 & 98.56 & 78.99 & 86.70 & 98.89 & 82.98 & 89.81 \\
\multicolumn{1}{l|}{\textbf{GOST}} & 18.66 & 99.78 & 93.84 & 96.53 & 5.34 & 99.63 & 98.1 & 98.81 & 1.51 & 99.57 & 99.4 & 99.48  & 91.12 & 72.95 & 75.84 & 85.09 & 88.08 & 83.93 & 82.81 & 93.49 & 86.60 \\
\multicolumn{1}{l|}{\textbf{aha-mont64}} & 16.69 & 100.0 & 97.68 & 98.8 & 1.38 & 100.0 & 99.8 & 99.90 & 0.19 & 100.0 & 99.97 & 99.99  & 100.0 & 67.10 & 73.69 & 100.0 & 95.20 & 96.80 & 100.0 & 98.5 & 99.18 \\
\multicolumn{1}{l|}{\textbf{crc32}} & 24.23 & 99.00 & 96.14 & 97.27 & 7.58 & 89.37 & 98.91 & 89.17 & 7.23 & 96.28 & 98.93 & 96.51  & 88.58 & 53.23 & 62.93 & 79.63 & 70.58 & 70.49 & 85.00 & 70.07 & 75.94 \\
\multicolumn{1}{l|}{\textbf{cubic}} & 20.02 & 100.0 & 97.21 & 98.57 & 12.06 & 100.0 & 98.29 & 99.13 & 6.563 & 100.0 & 99.05 & 99.52  & 99.56 & 85.99 & 91.72 & 99.72 & 90.85 & 94.77 & 99.72 & 94.52 & 96.93 \\
\multicolumn{1}{l|}{\textbf{edn}} & 12.21 & 99.87 & 98.27 & 99.05 & 4.38 & 99.75 & 99.36 & 99.55 & 3.25 & 99.72 & 99.52 & 99.62  & 99.15 & 78.13 & 86.30 & 99.03 & 89.89 & 93.9 & 99.27 & 91.89 & 95.29 \\
\multicolumn{1}{l|}{\textbf{huffbench}} & 20.76 & 99.00 & 96.62 & 97.51 & 9.38 & 95.35 & 98.66 & 96.29 & 6.48 & 99.48 & 99.06 & 99.26  & 99.00 & 85.01 & 90.67 & 90.00 & 87.48 & 85.92 & 94.12 & 94.06 & 92.01 \\
\multicolumn{1}{l|}{\textbf{matmult-int}} & 13.69 & 98.70 & 98.11 & 98.37 & 2.76 & 97.72 & 99.61 & 98.64 & 0.28 & 97.02 & 99.96 & 98.47  & 98.58 & 88.43 & 92.27 & 98.07 & 97.10 & 97.39 & 96.72 & 99.65 & 98.12 \\
\multicolumn{1}{l|}{\textbf{md5sum}} & 19.24 & 99.86 & 97.35 & 98.56 & 3.72 & 99.87 & 99.45 & 99.66 & 3.84 & 99.99 & 99.44 & 99.71  & 98.26 & 87.08 & 91.45 & 99.11 & 96.57 & 97.78 & 99.47 & 96.44 & 97.92 \\
\multicolumn{1}{l|}{\textbf{minver}} & 12.8 & 100.0 & 98.25 & 99.09 & 1.90 & 100.0 & 99.73 & 99.86 & 0.00 & 100.0 & 100.0 & 100.0  & 100.0 & 91.86 & 95.13 & 99.96 & 98.63 & 99.2 & 99.97 & 100.0 & 99.98 \\
\multicolumn{1}{l|}{\textbf{nettle-aes}} & 14.33 & 100.0 & 98.02 & 98.98 & 3.83 & 100.0 & 99.44 & 99.72 & 2.96 & 100.0 & 99.57 & 99.78  & 99.91 & 90.14 & 94.26 & 99.57 & 96.62 & 98.03 & 99.51 & 97.23 & 98.35 \\
\multicolumn{1}{l|}{\textbf{nsichneu}} & 12.77 & 97.29 & 98.21 & 97.65 & 5.33 & 93.06 & 99.21 & 95.93 & 2.90 & 89.9 & 99.53 & 94.43  & 100.0 & 51.30 & 64.80 & 100.0 & 65.01 & 77.12 & 100.0 & 68.90 & 81.41 \\
\multicolumn{1}{l|}{\textbf{picojpeg}} & 11.53 & 100.0 & 98.38 & 99.17 & 1.97 & 100.0 & 99.72 & 99.86 & 0.08 & 100.0 & 99.99 & 99.99  & 100.0 & 91.88 & 95.34 & 100.0 & 98.48 & 99.16 & 100.0 & 99.92 & 99.96 \\
\multicolumn{1}{l|}{\textbf{primecount}} & 21.05 & 72.00 & 96.05 & 82.27 & 8.07 & 72.00 & 98.41 & 83.15 & 3.69 & 72.00 & 99.26 & 83.46  & 53.00 & 77.99 & 62.10 & 53.00 & 88.89 & 66.14 & 53.00 & 94.19 & 67.76 \\
\multicolumn{1}{l|}{\textbf{qrduino}} & 21.98 & 98.43 & 96.45 & 97.03 & 7.43 & 95.21 & 98.93 & 95.65 & 2.57 & 92.42 & 99.60 & 95.16  & 98.00 & 83.45 & 89.26 & 92.00 & 87.99 & 88.76 & 78.00 & 75.94 & 76.95 \\
\multicolumn{1}{l|}{\textbf{sglib-combined}} & 13.56 & 100.0 & 98.15 & 99.04 & 0.53 & 100.0 & 99.92 & 99.96 & 0.00 & 100.0 & 100.0 & 100.0  & 100.0 & 91.24 & 94.80 & 100.0 & 99.57 & 99.77 & 100.0 & 100.0 & 100.0 \\
\multicolumn{1}{l|}{\textbf{slre}} & 24.57 & 100.0 & 96.62 & 98.25 & 10.65 & 100.0 & 98.48 & 99.23 & 8.06 & 100.0 & 98.84 & 99.41  & 100.0 & 83.66 & 90.41 & 100.0 & 91.62 & 95.42 & 100.0 & 93.23 & 96.40 \\
\multicolumn{1}{l|}{\textbf{statemate}} & 16.45 & 99.98 & 97.72 & 98.82 & 5.40 & 99.99 & 99.22 & 99.6 & 1.68 & 100.0 & 99.75 & 99.87  & 100.0 & 88.63 & 93.44 & 100.0 & 95.59 & 97.62 & 100.0 & 98.44 & 99.21 \\
\multicolumn{1}{l|}{\textbf{tarfind}} & 10.59 & 100.0 & 98.52 & 99.24 & 2.71 & 100.0 & 99.61 & 99.80 & 1.08 & 100.0 & 99.84 & 99.92  & 99.67 & 92.68 & 95.6 & 99.01 & 97.87 & 98.34 & 98.57 & 98.97 & 98.75 \\
\multicolumn{1}{l|}{\textbf{wikisort}} & 19.65 & 100.0 & 97.26 & 98.59 & 9.43 & 99.98 & 98.65 & 99.31 & 6.50 & 99.97 & 99.05 & 99.51 & 99.96 & 86.16 & 92.04 & 99.9 & 92.47 & 95.85 & 99.83 & 94.29 & 96.93
\end{tabular}
}
\label{tab:ROP_ablation}
\label{tab:DOP_ablation}
\end{table*}

\section{Runtime Overhead}\label{sec:eval_runtime}

We evaluate \ourname's runtime overhead generated by the system, in terms of storage and computation times for the phases of preprocessing, training and inference. In \Cref{tab:runtime} we report the results. We perform this analysis on a Raspberry Pi 4B 2GB. However, in a real-world scenario, the verifier is a server with high computational power. 

Note that the preprocessing phase can already take place while the software is running, as the graph can be built at runtime as soon as every execution step of the current execution is collected, drastically reducing the impact on the system. 
Moreover, from the preprocessing timings, it can be seen how this phase scales linearly to the trace length (cf., \Cref{subsec:data_pre}).

The evaluation shows that even on resource-constrained devices, the training and inference phase are extremely efficient (i.e., less than 5 minutes and less than 0.15 seconds, respectively) thanks to \ourname's lightweight model allowing a swift attestation. 
Further, storage-efficiency-wise dealing with structured data (i.e., graphs) allows the system to reduce the storage/communication overhead significantly, achieving a compression of up to 276.90 times compared to a raw trace.
Moreover, the runtime memory footprint shows the applicability of our approach to resource-constrained devices as the model occupies in average around 70.10KB, while the graph data around 0.73KB.

\begin{table}[H]
\centering
\caption{Runtime measurements for each dataset. We report the timings for preprocessing, training time and inference. Further, we report the compression obtained when changing the representation of a trace to graph. All the experiments are executed on a Raspberry Pi 4B 2GB.}
\label{tab:runtime}

\resizebox{.9\columnwidth}{!}{%
\begin{tabular}{c|ccccc}
\textbf{Dataset} & \textbf{Trace Size}  & \textbf{Graph Size} & \textbf{Preprocessing}& \textbf{Training time} & \textbf{Inference} \\
& \textbf{(MB)} & \textbf{(MB)(Compression)} & \textbf{(s)}& \textbf{(s)} & \textbf{(s)}\\\hline
DH & 2.46 & 0.24 (10.35$\times$) & 2.59 & 53.16 & 0.04 \\
AES & 27.86 & 0.89 (31.38$\times$) & 17.55 & 215.70 & 0.13 \\
GOST & 23.37 & 0.90 (26.05$\times$) & 14.99 & 352.50 & 0.10 \\
DES & 247.71 & 0.89 (276.90$\times$) & 149.00 & 245.00 & 0.13 \\
DESX & 252.77 & 0.92 (275.00$\times$) & 151.00 & 313.20 & 0.11 \\
aha-mont64 & 6.84 & 0.19 (36.05$\times$) & 4.00 & 31.59 & 0.06 \\
crc32 & 9.26 & 0.19 (49.64$\times$) & 5.22 & 51.97 & 0.05 \\
cubic & 2.75 & 0.23 (11.85$\times$) & 1.98 & 57.74 & 0.06 \\
edn & 6.60 & 0.20 (33.79$\times$) & 3.75 & 62.48 & 0.04 \\
huffbench & 8.81 & 0.20 (43.55$\times$) & 4.91 & 36.06 & 0.04 \\
matmult-int & 7.21 & 0.19 (37.49$\times$) & 4.06 & 35.37 & 0.06 \\
md5sum & 5.87 & 0.25 (23.82$\times$) & 3.70 & 67.57 & 0.06 \\
minver & 1.92 & 0.19 (9.96$\times$) & 1.42 & 101.00 & 0.06 \\
nettle-aes & 1.54 & 0.19 (8.00$\times$) & 1.17 & 68.18 & 0.06 \\
nsichneu & 13.52 & 0.26 (51.14$\times$) & 8.20 & 81.98 & 0.06 \\
picojpeg & 7.55 & 0.22 (33.86$\times$) & 4.49 & 62.49 & 0.05 \\
primecount & 30.53 & 0.19 (161.70$\times$) & 16.78 & 76.18 & 0.05 \\
qrduino & 8.12 & 0.23 (34.60$\times$) & 4.93 & 84.61 & 0.05 \\
sglib-combined & 11.29 & 0.22 (51.85$\times$) & 6.64 & 58.27 & 0.06 \\
slre & 10.94 & 0.20 (54.12$\times$) & 6.56 & 68.16 & 0.06 \\
statemate & 2.50 & 0.19 (12.88$\times$) & 1.79 & 55.22 & 0.04 \\
tarfind & 2.91 & 0.19 (15.33$\times$) & 1.87 & 36.64 & 0.05 \\
wikisort & 3.89 & 0.21 (18.77$\times$) & 2.42 & 37.30 & 0.06 \\
\end{tabular}%

}
\end{table}
\end{appendices}

\end{document}